\documentstyle[12pt]{article}
\newcommand{\extraspace}{\addtolength{\abovedisplayskip}{2mm}
\addtolength{\belowdisplayskip}{2mm}
\addtolength{\abovedisplayshortskip}{2mm}
\addtolength{\belowdisplayshortskip}{2mm}}
\newcommand{\be}{\begin{equation}\extraspace}

\newcommand{\ee}{\end{equation}}

\newcommand{\bea}{\begin{eqnarray}\extraspace}
\newcommand{\beastar}{\begin{eqnarray*}\extraspace}
\newcommand{\eea}{\end{eqnarray}}
\newcommand{\eeastar}{\end{eqnarray*}}

\newcommand{\nonu}{\nonumber \\[2mm]}

\newcommand{\stru}{\rule[-1mm]{0mm}{6mm}}

\newcommand{\half}{\frac{1}{2}}

\newcommand{\La}{\Lambda}

\newcommand{\eps}{\epsilon}

\newcommand{\la}{\lambda}

\newcommand{\cW}{{\cal W}}
\newcommand{\vac}{|0\rangle}

\newcommand{\mod}{{\rm mod}}

\def\lefthook{{\vrule height5pt width0.4pt depth0pt}}
\def\righthook{{\vrule height5pt width0.4pt depth0pt}}
\def\leftrighthookfill{$\mathsurround=0pt \mathord\lefthook
\hrulefill\mathord\righthook$}
\def\underhook#1{\vtop{\ialign{##\crcr$\hfil\displaystyle{#1}\hfil$\crcr
\noalign{\kern-1pt\nointerlineskip\vskip2pt}
\leftrighthookfill\crcr}}}

\newcommand{\newsection}[1]{
\vspace{12mm}
\pagebreak[3]
\addtocounter{section}{1}
\setcounter{equation}{0}
\setcounter{subsection}{0}
\setcounter{footnote}{0}
\begin{flushleft}
{\large\bf \thesection. #1}
\end{flushleft}
\nopagebreak}

\def\ZZ{Z\!\!\!Z} 		
\def\CC{I\!\!\!\!C}		%
\def\QQ{I\!\!\!\!Q}		

\def\cM{{\cal M}} \def\cO{{\cal O}}
\def\al{\alpha} \def\ze{\zeta}
\def\qbin#1#2{ \left[ \begin{array}{c} {#1} \\ {#2} \end{array} \right]}
\def\bin#1#2{ \left( \begin{array}{c} {#1} \\ {#2} \end{array} \right)}
\def\ccft{c_{\,\rm CFT}}
\def\eql{~=~}
\def\id{{1\!\!1}}
\def\tP{P}
\def\wK{\widetilde{K}} \def\wc{\widetilde{c}}

\expandafter\ifx\csname amssym.def\endcsname\relax \else\endinput\fi
\expandafter\edef\csname amssym.def\endcsname{%
       \catcode`\noexpand\@=\the\catcode`\@\space}
\catcode`\@=11
\def\undefine#1{\let#1\undefined}
\def\newsymbol#1#2#3#4#5{\let\next@\relax
 \ifnum#2=\@ne\let\next@\msafam@\else
 \ifnum#2=\tw@\let\next@\msbfam@\fi\fi
 \mathchardef#1="#3\next@#4#5}
\def\mathhexbox@#1#2#3{\relax
 \ifmmode\mathpalette{}{\m@th\mathchar"#1#2#3}%
 \else\leavevmode\hbox{$\m@th\mathchar"#1#2#3$}\fi}
\def\hexnumber@#1{\ifcase#1 0\or 1\or 2\or 3\or 4\or 5\or 6\or 7\or 8\or
 9\or A\or B\or C\or D\or E\or F\fi}

\font\tenmsa=msam10 scaled \magstep1
\font\sevenmsa=msam7 scaled \magstep1
\font\fivemsa=msam5 scaled \magstep1
\newfam\msafam
\textfont\msafam=\tenmsa
\scriptfont\msafam=\sevenmsa
\scriptscriptfont\msafam=\fivemsa
\edef\msafam@{\hexnumber@\msafam}
\mathchardef\dabar@"0\msafam@39
\def\dashrightarrow{\mathrel{\dabar@\dabar@\mathchar"0\msafam@4B}}
\def\dashleftarrow{\mathrel{\mathchar"0\msafam@4C\dabar@\dabar@}}

\def\ulcorner{\delimiter"4\msafam@70\msafam@70 }
\def\urcorner{\delimiter"5\msafam@71\msafam@71 }
\def\llcorner{\delimiter"4\msafam@78\msafam@78 }
\def\lrcorner{\delimiter"5\msafam@79\msafam@79 }
\def\yen{{\mathhexbox@\msafam@55}}
\def\checkmark{{\mathhexbox@\msafam@58}}
\def\circledR{{\mathhexbox@\msafam@72}}
\def\maltese{{\mathhexbox@\msafam@7A}}

\font\tenmsb=msbm10 scaled \magstep1
\font\sevenmsb=msbm7 scaled \magstep1
\font\fivemsb=msbm5 scaled \magstep1
\font\tamsb=msbm10 scaled \magstep2
\newfam\msbfam
\textfont\msbfam=\tenmsb
\scriptfont\msbfam=\sevenmsb
\scriptscriptfont\msbfam=\fivemsb
\edef\msbfam@{\hexnumber@\msbfam}
\def\Bbb#1{{\fam\msbfam\relax#1}}
\def\widehat#1{\setbox\z@\hbox{$\m@th#1$}%
 \ifdim\wd\z@>\tw@ em\mathaccent"0\msbfam@5B{#1}%
 \else\mathaccent"0362{#1}\fi}
\def\widetilde#1{\setbox\z@\hbox{$\m@th#1$}%
 \ifdim\wd\z@>\tw@ em\mathaccent"0\msbfam@5D{#1}%
 \else\mathaccent"0365{#1}\fi}
\font\teneufm=eufm10 scaled \magstep1
\font\seveneufm=eufm7 scaled \magstep1
\font\fiveeufm=eufm5 scaled \magstep1
\font\taeufm = eufm10 scaled\magstep2

\newfam\eufmfam
\textfont\eufmfam=\teneufm
\scriptfont\eufmfam=\seveneufm
\scriptscriptfont\eufmfam=\fiveeufm
\def\frak#1{{\fam\eufmfam\relax#1}}

\font\teneufb = eufb10 scaled \magstep1

\newfam\eufbfam
\textfont\eufbfam=\teneufb

\font\tencmbsy  = cmbsy10 scaled \magstep1

\font\sevencmbsy = cmbsy7 scaled \magstep1
\font\fivecmbsy = cmbsy5 scaled \magstep2

\newfam\cmbsyfam
\textfont\cmbsyfam=\tencmbsy
\scriptfont\cmbsyfam=\sevencmbsy
\scriptscriptfont\cmbsyfam=\fivecmbsy

\csname amssym.def\endcsname

\def\ZZ{{\Bbb Z}}
\def\sZZ{{\tamsb \ZZ}}

\def\CC{{\Bbb C}}

\def\fg{{\frak g}} \def\bfg{\fg}
\def\fso{{\frak{so}}}
\def\fsl{{\frak{sl}}}
\def\fsu{{\frak{su}}}
\def\fsp{{\frak{sp}}}

\def\sfso{{\taeufm \fso}}

\def\sfsu{{\taeufm \fsu}}
\def\sfsp{{\taeufm \fsp}}

\def\Journal#1#2#3#4{{#1} {{#2}} ({#3}) {#4}}
\def\CMP{Comm.\ Math.\ Phys.\ }
\def\NPB{Nucl.\ Phys.\ B}
\def\PRL{Phys.\ Rev.\ Lett.\ }
\def\PRB{Phys.\ Rev.\ B}
\def\PLA{Phys.\ Lett.\ A}
\def\PLB{Phys.\ Lett.\ B}
\def\JSM{J.\ Sov.\ Math.\ }
\def\SMD{Sov.\ Math.\ Dokl.\ }
\def\JPA{J.\ Phys.\ A}
\def\JPAA{J.\ Pure Appl.\ Algebra\ }
\def\MPLA{Mod.\ Phys.\ Lett.\ A}
\def\MPLB{Mod.\ Phys.\ Lett.\ B}
\def\DMJ{Duke Math.\ J.\ }
\def\LMP{Lett.\ Math.\ Phys.\ }
\def\PTPS{Prog.\ Theor.\ Phys.\ Suppl.\ }
\def\NCim{Nuovo Cimento}
\def\JSP{J.\ Stat.\ Phys.\ }
\def\JETP{Sov.\ Phys.\ JETP}
\def\PAWPM{S.-B Preuss.\ Akad.\ Wiss.\ Phys.\ Math.\ Kl.\ }
\def\DMath{Discr.\ Math.\ }
\def\SJMA{SIAM J.\ Math.\ Anal.\ }
\def\JAMS{J.\ Austral.\ Math.\ Soc.\ (Series A)\ }

\def\IJMPA{Int.\ J.\ Mod.\ Phys.\ A}
\def\IJMPB{Int.\ J.\ Mod.\ Phys.\ B}
\def\JMP{J.\ Math.\ Phys.\ }
\def\PLMS{Proc.\ London Math.\ Soc.\ }
\def\SMNS{Selecta Math.\ (N.S.)\ }

\begin{document}
\baselineskip=17pt

\vskip 1.5cm
\begin{center}

{\Large Exclusion Statistics in Conformal Field Theory}\\
\vskip 4mm
{\large -- generalized fermions and spinons for level-1 WZW theories --}
\vskip 1.5cm
{\large Peter Bouwknegt}
\vskip .3cm
{\sl Department of Physics and Mathematical Physics \\
University of Adelaide \\
Adelaide, SA 5005, AUSTRALIA}

\vskip .9cm
{\large Kareljan Schoutens}
\vskip .3cm
{\sl Institute for Theoretical Physics \\
University of Amsterdam \\
Valckenierstraat 65 \\
1018 XE  Amsterdam, THE NETHERLANDS}

\vskip .7cm
{\bf Abstract}
\end{center}
\baselineskip=15pt
{\small
We systematically study the exclusion statistics for quasi-particles for 
Conformal Field Theory spectra by employing a method based on 
recursion relations for truncated spectra. Our examples include generalized 
fermions in $\ccft<1$ unitary minimal models, $\ZZ_k$ parafermions, and 
spinons for the $\fsu(n)_1$, $\fso(n)_1$ and 
$\fsp(2n)_1$ Wess-Zumino-Witten models. For some of the latter 
examples we present explicit expressions for finitized affine characters 
and for the \hbox{$N$-spinon} decomposition of affine characters.
}
  
\vfill
\leftline{ADP-98-13/M65}
\noindent{ITFA-98-11 \hfill October 1998} \newline
\noindent{{\tt hep-th/9810113} \hfill  revised March 1999}

\newpage

\baselineskip=17pt
\newsection{Introduction}
\vskip 2mm

Among the most spectacular features of quantum many body 
systems in low dimensions are phenomena such as spin-charge separation 
and quantum number fractionalization. These terms refer to situations 
where fundamental excitations over a many-body ground state carry 
quantum numbers (for spin, charge, etc.) which are fractions of the 
quantum numbers carried by the microscopic degrees of freedom in the 
system. Examples are fractionally charged (Laughlin) 
quasi-particles in fractional quantum Hall systems, spinons
in antiferromagnetic spin-chains and the spinons and holons 
for $d=1$ itinerant electrons with repulsive interactions. 
Recent experiments on fractional quantum Hall systems and on spin-chain
compounds have directly probed various aspects of these
`fractional quasi-particles'. It has been suggested that the physics 
of quantum critical points in dimension higher than one may involve the 
same phenomenon of quantum number fractionalization \cite{La}.

An essential aspect of quasi-particles carrying unusual 
(fractionalized) quantum numbers is that their statistics will 
be equally unusual. Depending on the context, one may wish to 
consider braid statistics (in 2 spatial dimensions) or exclusion 
statistics, the two notions being closely related. Prototypical 
examples are quasi-hole excitations 
over the Laughlin ground states for the $\nu={1 \over m}$ fractional 
quantum Hall effect. Such quasi-holes satisfy fractional (anyonic) 
braid statistics \cite{HASW}, and obey a specific form of fractional 
exclusion statistics (see below).

The term `fractional exclusion statistics' was introduced in Haldane's
1991 paper \cite{Ha2}, which proposed a particular generalization
of the Pauli principle. This definition leads to the notion
of a `quantum gas of non-interacting particles satisfying fractional
exclusion statistics'. The thermodynamic properties of such gases
have been analyzed in, e.g., \cite{gstats}.

There exist a number of different approaches by which the
`fractional exclusion statistics' (Haldane's or more general)
of (quasi-)particles in specific quantum many body systems
can be investigated.

In models that are solvable by the Bethe Ansatz, the Bethe equations 
for a set of fundamental rapidities can sometimes be interpreted in
terms of exclusion statistics. This approach has been followed
for the $SU(2)$ Haldane-Shastry spin-chain \cite{Ha1} 
and for a specific integrable $3$-state Potts chain \cite{3-potts}. 
The character expressions that resulted from the latter example
have been generalized to a large category of `fermionic sum
formulas' for the characters in a variety of models
of Conformal Field Theory (see, e.g., 
\cite{ferchar1,ferchar2,FQOW,ferchar3,FS,Geo,HKKOTY,Suz} and 
references therein).

If scattering ($S$-matrix) data for the quasi-particles of
choice are known, one may employ the Thermodynamic Bethe Ansatz 
(TBA) to determine the corresponding exclusion statistics.
For a particular class of $S$-matrices, the TBA statistics agree
with `fractional exclusion statistics' in the sense of Haldane
\cite{TBA}. A spectacular application of this TBA approach
has been the exact computation of a universal conductance
curve for edge-to-edge tunneling of quasi-particles in the
fractional quantum Hall effect (FQHE) \cite{FLS}.
 
A third and independent approach to `fractional exclusion
statistics' is possible for systems that are described by
an (effective) Conformal Field Theory (CFT). This approach
will be central to the work presented in this paper.
In general, a (rational) CFT comes with a precise list of 
primary field operators from which one may extract creation 
and annihilation operators for various quasi-particles. 
In this manner, many 
important examples of quantum number fractionalization are 
elegantly described. In addition, the exclusion statistics 
properties of quasi-particles can conveniently be studied in 
the CFT setting, by following a method proposed in \cite{Sc1}. 
This method, which employs 
recursion relations for truncations (`finitizations') of the 
chiral CFT spectrum, leads to explicit expressions for
single-level partition sums and hence exposes the underlying
exclusion statistics. We stress that this approach does 
not rely on underlying Bethe equations and/or scattering data 
for the quasi-particles, but is instead directly based on 
algebraic properties of CFT (primary) fields. This observation 
is important as in many cases of interest Bethe equations or 
scattering data are simply not available.

In \cite{ES,Sc2}, the recursion method was applied to CFT's 
describing edge theories for a variety of FQHE states and used
to determine 
the exclusion statistics of edge quasi-particles. Of particular 
interest are the so-called pfaffian QHE states, where the edge 
quasi-holes satisfy what is called non-abelian exclusion statistics
\cite{Sc2}. In \cite{ES}, it was demonstrated how the exclusion 
statistics properties of edge quasi-particles manifest themselves in 
equilibrium properties (notably: the Hall conductance) and in
transport properties.

To further illustrate the `CFT approach' to quasi-particle 
statistics, we recall that the excitation spectrum for 
critical spin-$S$ spin 
chains (with $S \geq 1$) can be understood in terms of fundamental 
spinons carrying spin-$\half$. This remarkable result, which can be 
derived from a Bethe Anstaz exact solution \cite{Ta}, 
is immediately clear when one recognizes that the effective CFT 
is a $\fsu(2)_{k=2S}$ Wess-Zumino-Witten (WZW) model, 
which has a $j=\half$ primary 
`spinon' field in its spectrum. A spinon basis for the $\fsu(2)_k$
spectrum was presented in \cite{BLS2} and the corresponding 
spinon statistics were described in \cite{FrS}.

In some special cases, the exclusion statistics of CFT quasi-particles
(as obtained from the recursion method) turn out to agree with
`fractional exclusion statistics' as defined by Haldane.  In such
cases, the quasi-particle character formulas assume the form of
`fermionic sum formulas'.  Comparing these with the results of the
Stony Brook group \cite{ferchar1,ferchar2}, one concludes that in
these special cases an analysis using `quasi-particles associated to
conformal fields' and an approach based on `Bethe Ansatz
quasi-particles' lead to identical results. The correspondence between
`fermionic sum formulas' and Haldane's statistics was first made in a
large number of special cases (see \cite{Hik,BLS1,ES,Ga}) and was
recently put on a general footing in \cite{BM} (see also
\cite{Scomm}).  We would like to stress that the most general
`fermionic sum formulas' (see Eq.~(\ref{eqBs})), with finite values
for some of the parameters $u_a$, correspond to a form of non-abelian
exclusion statistics.  (An example is provided by the abovementioned
$\fsu(2)_k$ spinons, see \cite{BLS2,FrS}.)  For many of the examples
that we treat in this paper, notably the spinons for level-1 WZW
models, the quasi-particle character formulas are not of the
`fermionic sum' type and appear to be more general than what has been
considered in the literature.

In the present paper we present a systematic study of some
of the more interesting examples of exclusion statistics for CFT 
quasi-particles. Throughout this paper, we shall be employing
the recursion method of \cite{Sc1} to determine the exclusion
statistics and the associated thermodynamics of specific CFT 
quasi-particles. In section 2.1 we review this method, paying 
special attention to the possibility of non-abelian exclusion 
statistics. In section 2.2 we briefly review Haldane's approach,
and we derive equations that determine the CFT central charge
for a given choice of statistics matrix. In section 2.3 we
treat two proto-typical examples, which are the Majorana fermion 
and $\fsu(2)_1$ spinons, and in section 2.4 we discuss in general 
terms character expressions that are associated to a CFT 
quasi-particle basis. In section 3 we discuss generalized fermions
in $c_{\rm CFT}<1$ minimal models and section 4 is devoted to $\ZZ_k$
parafermions. In section 5 we discuss spinons for $\fsu(n)_1$ 
WZW models. Section 6 is devoted to $\fso(n)_1$ WZW models,
which we analyze in terms of quasi-particles transforming in the
spinor representation(s) of $\fso(n)$. In section~7 we briefly discuss 
the case $\fsp(2n)_1$ WZW. For some of the examples
that we treat, we provide explicit character formulas, both for the 
finitized characters and the $N$-particle truncated characters.
Section 8 contains some further remarks on the WZW theories,
and in section 9 we offer a brief outlook.

Possible physical applications of the results and 
approach discussed in this paper have been discussed in, e.g.,
\cite{ES,Sc2,FrS,BS2}.

\newsection{General structure}
\subsection{Introducing the method}

We start by introducing in general terms the `recursion method',
which will be applied throughout this paper. 

The subject of study is what is called the chiral spectrum
or chiral Hilbert space of a conformal field theory. In a 
Rational Conformal Field Theory the chiral spectrum consists
of a finite number of
irreducible modules of the Chiral Algebra. (To be precise,
these will be the modules that participate in the modular invariant
partition function.) Depending on the choice of theory, the
Chiral Algebra can be, e.g., the Virasoro algebra, a $\cW$-type
extended algebra or an affine Lie algebra. We shall
encounter examples of all three possibilities in this paper.

By a quasi-particle approach towards a RCFT we mean 
a formulation where we interpret the chiral spectrum as
a collection of states, each of which is generated by the 
repeated action of creation operators for CFT quasi-particles
on a suitable, finite, set of reference states. The creation 
operators are nothing else than the Fourier modes of a selected 
set of (primary) field operators. Denoting these modes by
$\phi^{(a)}_{-s}$, we are thus considering states of the type
\be
\phi^{(a_N)}_{-s_N} \ldots\phi^{(a_2)}_{-s_2}\phi^{(a_1)}_{-s_1}
| 0 \rangle_J
\label{eqBa}
\ee
In this notation, the label $a$ enumerates the selected 
primary fields and the index $J$ labels the various reference 
states.

For a quasi-particle formulation to be complete we require that the 
collection of states (\ref{eqBa}) spans the complete chiral spectrum.
When this condition is met, one likes to reduce the collection 
(\ref{eqBa}) in such a manner that the reduced set precisely 
forms a basis for the chiral Hilbert space. In first approximation, 
the restriction will amount to ordering the $s_i$, $i=1,2,\ldots$, 
in ascending order, the precise details depending on the case at hand.
Obviously, the prototype for constructions of this kind is a theory 
of free fermions, where the canonical anti-commutation relations of 
the fermion fields determine the systematics. 
For quasi-particles satisfying abelian braiding statistics, one may 
derive `generalized commutation relations', which may then be employed 
in the reduction of (\ref{eqBa}) to a basis set.  The $\fsu(2)_1$ WZW 
theory provides an example where this procedure has been carried out 
in explicit detail \cite{BLS1}. In theories with non-abelian braiding, 
simple algebraic relations do not seem to be available and a systematic 
procedure for obtaining a quasi-particle basis is presently not known
(except after $q$-deformation in the crystal limit, i.e., the $q\to0$ 
limit \cite{NYa}). 
Examples of non-abelian theories where explicit quasi-particle bases 
have been proposed are the $\fsu(2)_k$ WZW theories \cite{BLS2}, and the 
CFT for the so-called pfaffian quantum Hall state \cite{Sc2}.

Having obtained an explicit quasi-particle basis for a given chiral 
spectrum, one may go ahead and try to translate the result into a 
statement about the exclusion statistics of the quasi-particles. 
In \cite{Sc1}, one of us proposed 
the following procedure: one starts by restricting a basis of 
quasi-particle states
of the form (\ref{eqBa}) by requiring that the participating momenta
(modes) satisfy $s \leq l$.\footnote{In some cases,
such as $\fso(n)_1$, we allow for an
extra shift in $l$ of order one if this makes the recursion look simpler.}
One then defines truncated partition sums
as expressions of the type
\be 
P^{(\alpha)}_l(x_a,q) \eql
{\rm tr}_l^{\alpha} \left(  q^{L_0} \, \prod_a x_a^{N_a} \right) \,,
\label{eqBb}
\ee
where $x_a=e^{\beta \mu_a}$ denotes the fugacity of the particle 
of species $a$ and 
the superscript $\alpha$ denotes the restriction that
the full quasi-particle state belongs to a sector specified by $\alpha$. 
Typically, the label $\alpha$ will run over the irreducible 
representations of the Chiral Algebra.
Having defined the truncated partition sums $P^{(\alpha)}_l(x_a,q)$,
one is interested in their rate of growth,
\be
P^{(\alpha)}_{l+1}(x_a,q) / P^{(\alpha)}_l(x_a,q)  
~\sim~ \lambda(z_a=x_a q^l) 
\label{eqBc}
\ee
where, typically, $\lambda(z_a)$ will not depend on the sector $\alpha$. 
In practice, the functions $\lambda(z_a)$ can be obtained from 
recursion relations satisfied by the $P^{(\alpha)}_l$.  In fact,
for this purpose it suffices to consider the recursion relations
at the point $q=1$.
We can identify $\la(z_a=e^{\beta(\mu_a- \eps)})$ with the
grand partition function for a single energy level, hence
from $\lambda(z_a)$ one derives thermodynamic quantities
such as the appropriate generalization of the Fermi-Dirac distribution
function
\be
n_a(\eps) \eql 
z_a {\partial \over \partial z_a} \log \lambda(z)  _{\big|
z_b = e^{\beta(\mu_b-\eps)}}  \,.
\label{eqBd}
\ee

It is an easy exercise to express the specific heat $C$
of the CFT in terms of $\lambda(z_a)$. In the 
1-component case, the result is 
\be
C \eql \gamma \rho k_B^2 T  \ , \qquad
\gamma \eql 2 \int_0^1 dy \ {1 \over y} \log \lambda(y) \,,
\label{eqBe}
\ee
with $\rho$ the density of states, 
and comparing with the well-known result in terms of
the central charge $\ccft$, 
\be
\gamma \eql {\pi^2 \over 3} \, \ccft\,, \label{eqBf}
\ee
one obtains identities, which in concrete cases
lead to new proofs for certain di-logarithm identities
(see \cite{NRT,FSz} and references therein). 
In many cases, the central charge identity 
serves as a first check on a conjectured quasi-particle basis
and/or recursion relation.

We now focus on some 
interesting limits of the expressions $\lambda(z_a)$.
A first limit of interest is $z_a \ll 1$, where
one may expand
\be
\lambda(z_a) \eql 1 + \sum_b \alpha_b z_b + {\cal O}(z^2)\,.
\label{eqBg}
\ee
The quantities $\alpha_a$ manifest themselves as prefactors in
the Boltzmann tails
\be
n_a(\eps) ~\stackrel{\eps \to \infty}{\sim}~ \alpha_a \, 
  e^{-\beta \eps}\,,
\label{eqBh}
\ee
of the generalized distribution functions. As a general result,
a factor $\alpha_a \neq 1$ signals non-abelian braiding
statistics of the associated CFT primary fields $\phi^a(z)$.
A precise statement is that $\alpha_a$ equals the largest eigenvalue
of the incidence matrix associated to the fusion rules of the
primary field $\phi^a(z)$ with the other primary fields
in the theory.  This can be seen as follows.  Suppose the chiral algebra 
has a set of fusion rules
\be 
a\times b \eql \sum_c\ N_{ab}{}^c \, c\,, \label{eqBHa}
\ee
then the recursion relations for the truncated
characters $P^{(i)}_l(x_a,q)$ (at $q=1$) will be of the form 
\be
P^{(i)}_{l+1} \eql P_l^{(i)} + \sum_{a,j} x_a N_{ai}{}^j P^{(j)}_{l+1-a_j}
  + \cO(x^2) \,,\label{eqBhb}
\ee
for some set $a_j\in\QQ_{\geq0}$.  Asymptotically, $P_l^{(i)} \sim
\mu_i(x) \la(x)^l$.  Substituting this in (\ref{eqBhb}) gives
\be
\mu_i(x) \la(x)^l (\la(x) -1) \eql \sum_{a,j} x_a 
N_{ai}{}^j \mu_j(x) \la(x)^{l+1-a_j} + \cO(x^2) \,.
\label{eqBhc}
\ee
Expanding $\mu_i(x) = \mu_i + \cO(x)$ and $\la(x) = 1 + \sum_a \al_a x_a
+\cO(x^2)$ as in (\ref{eqBg}), we find the following equation for the 
$\cO(x_a)$ term
\be
\sum_j N_{ai}{}^j \mu_j \eql \al_a \mu_i\,, \label{eqBhd}
\ee
which proves that $\al_a$ is the largest (real) eigenvalue (and
$\mu_i$ the eigenvector) of the 
fusion matrix $(N_a)_i{}^j = N_{ai}{}^j$ corresponding to the primary 
field $\phi_a(z)$.
Below we encounter concrete examples of this statement.

A second interesting limit is $z_a \to \infty$ in some fixed
ratio. For example, setting all $z_{a'}$ with $a' \neq a$ equal 
to zero and keeping just $z_a$, one typically finds
$\lambda \sim z_a^{\beta_a}$. The number $\beta_a$ then represents the 
maximum $n_a^{\rm max}$ of the distribution function for particles of 
species $a$ in the absence of any others.

In the example of level-1 WZW models, the set of quasi-particles 
transform in an irreducible finite dimensional representation of
the underlying Lie algebra
(or two irreducibles in the case of $\fso(2n)$).  In that case we will
often associate the same fugacity $x_a=x$ to all particles in the 
representation and derive an equation for the total grand partition
function $\la_{\rm tot}(x) = \prod_a \la_a(x)$.  
The small $x$ expansion of $\la_{\rm tot}(x)$ 
is then given by $\la_{\rm tot}(x) = 1 + D \alpha x +\cO(x^2)$, where
$D$ is the dimension of the irreducible representation and $\alpha$ the
largest eigenvalue of the fusion matrix, while for the large $x$ 
behavior one typically finds $\la_{\rm tot} \sim x^\beta$.  We define 
$n^{\rm max}_{\rm tot}$ as this value of $\beta$.  The large $x$ 
behavior of the average partition function $\la_{\rm av}(x) = 
\la_{\rm tot}(x)^{1/D}$ defines similarly the number $n^{\rm max}_{\rm av} = 
n^{\rm max}_{\rm tot}/D$.

\subsection{Haldane's exclusion principle}

We conclude this general introduction with a brief review of 
Haldane's notion of (abelian) fractional 
exclusion statistics \cite{Ha2}.  It is based on the idea that 
the number of accessible states $d_a$ for a particle of species $a$
depends on the particle numbers $N_b$ of all the other particles 
through a statistical interaction matrix $G_{ab}$ by
\be 
{\partial d_a\over \partial N_b} \eql - G_{ab} \,.\label{eqBi}
\ee
For a `generalized ideal gas of fractional statistics particles'
this leads the following equations for the one-particle grand canonical 
partition functions $\lambda_a(z)$ (\cite{gstats}, see the last paper of this
reference for an explicit discussion of the multi-component case)
\be
\left( { \la_a-1\over \la_a}\right) \prod_b \la_b^{G_{ab}} \eql z_a \,,
\label{eqBj}
\ee
from which the 1-particle distribution functions can be recovered 
by eqn.\ (\ref{eqBd}). The system (\ref{eqBj}) always leads to a 
small $x$ expansion $\la_a(x) = 1 +  x_a +\cO(x^2)$, i.e., 
corresponds to abelian exclusion statistics.  

Starting from the eqns.\ (\ref{eqBe}) and (\ref{eqBf}), one may  
compute the central charge of a generalized ideal gas satisfying
Haldane exclusion statistics 
\be
\left( {\pi^2 \over 6} \right) \ccft \eql
\sum_a \ \int_0^1 {dz_a \over z_a}  \ \log\,\la_a(z_a)\,.
\label{eqBk}
\ee
{}From (\ref{eqBj}) we find, for each $a$,
\be
{d\la_a \over \la_a (\la_a-1)} + \sum_b G_{ab} {d\la_b\over \la_b} \eql 
{dz_a\over z_a} \,.  \label{eqBl}
\ee
Thus 
\be 
\left( {\pi^2 \over 6} \right) \ccft \eql 
\sum_a \int_{y_a}^1 {d\la_a\over \la_a(\la_a-1)} \log \la_a
+ \sum_{a,b} G_{ab} \int_{y_b}^1 {d\la_b\over \la_b }\log \la_a \,,
\label{eqBm}
\ee
where the $y_a= \la_a(z_a=1)$ are determined as a solution of 
(\ref{eqBj}) with all $z_a=1$.
Substituting (\ref{eqBj}) again, we find 
\be
\left( {\pi^2 \over 6} \right) \ccft 
\eql \sum_a \int_{y_a}^1 \left(
{d\la_a\over \la_a-1} \log \la_a -  {d\la_a\over\la_a} \log(\la_a-1)
\right) + \sum_a \int_{y_a}^1 {d\la_a\over \la_a} \log z_a \,.\label{eqBn}
\ee
The last term on the right hand side of (\ref{eqBn}) equals the left 
hand side (upto a sign) by partial integration.  Changing variables
to $\mu_a = (\la_a-1)/\la_a$ in the remaining terms finally gives
\be
\left( {\pi^2 \over 6} \right) \ccft \eql \sum_a \ L(x_a)\,,
\label{eqBo}
\ee
where 
\bea
L(x) & \eql & -{\textstyle{1\over2}} \int_0^x \ dy \left( 
{\log y \over 1-y } + {\log(1-y) \over y} \right) 
\eql {\rm Li}_2(x) + {\textstyle{1\over2}} \log x \log(1-x)\,,\nonu
{\rm Li_2}(x) & \eql & - \int_0^x \ dy\ {dy\over y}\,\log(1-y) \,, 
\label{eqBp}
\eea
are Rogers' and Euler's dilogarithm functions, respectively \cite{Lew}, 
and the $x_a$ are a solution of 
the TBA system
\be
x_a \eql \prod_b\, (1-x_b)^{G_{ab}} \,. \label{eqBq}
\ee
In the context of Haldane's statistics, the result (\ref{eqBo}), 
(\ref{eqBq}) was first given in \cite{BF}, where a slightly different 
derivation was presented.

The eqn.\ (\ref{eqBo}) is the same as the equation that 
determines the central charge of the quasi-particle character (see, 
e.g., \cite{RS,NRT,Kir})
\be
{\rm ch}(x_a,q) \eql
\sum_{n_a \atop {\rm relations} } \ \left(  \prod_a x_a^{n_a} \right)
\ {q^{{1\over2}
n_a G_{ab} n_b + C_a n_a} \over \prod_a (q)_{n_a} } \,. \label{eqBr}
\ee
Indeed, it has been conjectured \cite{BM} (see also \cite{Scomm}) that 
the quasi-particle character of a generalized ideal gas of particles 
with abelian exclusion statistics is precisely of the type (\ref{eqBr}). 
This connection has been established in \cite{Hik} for $g$-ons 
(i.e., the 1 component case of (\ref{eqBj})), and it has been
observed in several multi-component cases \cite{BLS1,Sc1,ES,Ga}.

The character (\ref{eqBr}) is the $u\to \infty$ limit of a more 
general `Universal Chiral Partition Function' \cite{BM}
\be 
{\rm ch}(x_a,q) \eql
\sum_{n_a \atop {\rm relations} } \ q^{{1\over2}
n_a G_{ab} n_b + C_a n_a} \left( \prod_a x_a^{n_a}\ 
\qbin{((1-{\bf G}){\bf n} +
 { {\bf u}\over2} )_a}{n_a}\right) \,, \label{eqBs}
\ee
with 
\be
\left[ \begin{array}{c} m \\ n \end{array} \right] \eql
 {(q)_m \over (q)_{m-n} (q)_n} \,, \qquad
(q)_n \eql \prod_{k=1}^n (1-q^k) \,. \label{eqBCb}
\ee
The exclusion statistics of  CFT quasi-particles that correspond to a
character formula (\ref{eqBs}) with some of the $u_a$ finite, are more general
than Haldane's. The CFT quasi-particles are associated to those values
$a=A$ for which $u_A=\infty$, while the other values $a=i$, with
$u_i<\infty$, correspond to auxiliary or pseudo-particles. After 
eliminating the auxiliary quantities $\lambda_i$ from the thermodynamic 
equations, one finds that $\lambda_A(x)=1+\alpha_A x_A + 
{\cal O}(x^2)$ with $\alpha_A \neq 1$, showing that the CFT 
quasi-particles obey non-abelian exclusion statistics. Fundamental examples 
are spinons in $k>1$ $\frak{su}(2)_k$ WZW theories \cite{BLS2,FrS}
and the generalized 
fermions that we discuss in Section 3. We refer to a forthcoming publication 
\cite{GS} for a detailed discussion.

It should be emphasized that not all examples of exclusion statistics
are most naturally described by characters of the form (\ref{eqBs}).  
Examples
are the $\frak{su}(n)_1$ WZW models which are based on
`Gentile parastatistics' rather than fermionic statistics \cite{Sc1}.
We refer to Section 5 for a discussion.

\subsection{Prototypes: free fermions and $\sfsu(2)_1$ spinons}

To get started we review the (trivial) example of the free Majorana 
fermion and the case of $\fsu(2)_1$ spinons, which we developed elsewhere 
\cite{Sc1}.
The CFT for a chiral Majorana fermion $\psi(z)$ with Neveu-Schwarz boundary 
conditions has two Virasoro sectors, with leading conformal dimensions 
$h_{(1,1)}=0$,  $h_{(2,1)}=\half$. We write the truncated characters as 
$P_l^1(x,q) \equiv P_l^{(1,1)}(x,q)$ and
$P_{l+{1 \over 2}}^2(x,q) \equiv P_{l+{1 \over 2}}^{(2,1)}(x,q)$
($l=0,1,\ldots$). 
In the limit $l \to \infty$, these reproduce the corresponding 
characters of the
$\ccft=\half$ Virasoro algebra. If we realize that the two sectors are 
generated by 
modes $\psi_{-l-\half}$,  with $l=0,1,\ldots$, satisfying canonical 
anti-commutation
relations, we immediately obtain the recursion relation
\be
\left( \begin{array}{c} P^1_{l+1} \\  P^2_{l+\half} \end{array} \right)
\eql \left( \begin{array}{cc} 
1	&  x  q^{l+\half} \\ x q^{l+\half} & 1 \end{array} \right)
\left( \begin{array}{c} P^1_{l} \\  P^2_{l-\half}  \end{array} \right) 
\,, \qquad l=0,1,2,\ldots\,.
\label{eqBBa}
\ee
Clearly, the associated one-particle partition function $\la(z)$ (cf.\
(\ref{eqBc})) is in this case simply given by the largest eigenvalue
of the recursion matrix in (\ref{eqBBa}), i.e., $\lambda(z)=1+z$,
where $z=x q^{l+{1\over2}}$, and the usual Fermi-Dirac distribution is
obtained.

The example of spinons for the $\fsu(2)_1$ theory has been treated 
elsewhere. In \cite{Ha1,HHTBP,BPS,BLS1} the spinon basis for the 
affine modules was discussed 
in great detail, and in \cite{Sc1} the systematics of this basis were 
translated into a recursion relation for the truncated characters 
$P^{(2j)}_l(x,t,q)$ with the 
$\fsu(2)$ spin taking the values $j=0,\half$ appropriate for the $k=1$ WZW 
model. It should be stressed that in the limit $l \to \infty$, these 
characters 
reproduce characters of the affine Lie algebra $\fsu(2)_1$. In a general 
RCFT, one expects characters of the Chiral Algebra of that theory. With the 
notation $\chi_{{\bf 2j+1}}$ for the $\fsu(2)$ characters, e.g.,
$\chi_{{\bf 2}}(t)=t+t^{-1}$ and $\chi_{{\bf 3}}(t)=t^2+1+t^{-2}$,  
the recursion relations take the form
\be
\left( \begin{array}{c} P^{(0)}_{l+1} \\ P^{(1)}_{l+1} \end{array} \right)
\eql \left( \begin{array}{cc} 
1+x^2 q^{2l+1} \, \chi_{\bf 3} &  x q^{l+{3 \over 4}}(1-x^2q^{2l}) \, 
\chi_{\bf 2} 
\\
x q^{l+{1 \over 4}} \, \chi_{\bf 2} & 1-x^2 q^{2l}
\end{array} \right)
\left( \begin{array}{c} P^{(0)}_{l} \\ P^{(1)}_{l} \end{array} \right) \,.
\label{eqBBb}
\ee
It is interesting to note that the subtractions that are part of the 
recursion have their origin in the symmetrization prescription which is 
part of the generalized Pauli Principle for spinons \cite{BLS1}. It has 
been demonstrated \cite{Sc1} (see also \cite{Ga}) that the thermodynamic 
distribution functions that follow from this recursion relation are 
identical to those associated to fractional exclusion statistics in 
the sense of Haldane, with statistical interaction matrix 
$G=\left( \begin{array}{cc} \scriptstyle{1\over2} & \scriptstyle{1\over2} \\
\scriptstyle{1\over2} & \scriptstyle{1\over2} \end{array}\right)$.

\subsection{Character identities}

The procedure outlined in section 2.1 relies on recursion
relations for truncated characters $P^{(\alpha)}_l(x_a,q)$,
and does not need closed form results for these characters.
Nevertheless, it is interesting and often illuminating
to consider exact character formulas, both for truncated
partition sums $P^{(\alpha)}_l(x_a,q)$ and their $l\to\infty$
limit, ${\rm ch}^{(\al)}(x_a,q)$, i.e., 
the character of the full CFT Hilbert space.  We have
\be 
{\rm ch}^{(\al)}(x_a,q) \eql \sum_{N_a\atop {\rm restrictions}} \ 
 \left( \prod_a \, x_a^{N_a} \right) \ {\rm ch}^{(N_a)}(q) \,,
\label{eqBCaa}
\ee
where ${\rm ch}^{(N_a)}(q)$ is referred to as the $N$-particle cut
of the quasi-particle basis (\ref{eqBa}).  In the explicit examples
of this paper we will often see that there is a remarkable `duality'
(involving $q\to q^{-1}$) between the truncated characters 
$P^{(\alpha)}_l(x_a=1,q)$ and the $N$-particle cuts ${\rm ch}^{(N_a)}(q)$.

For the example of the Majorana fermion, the truncated partition 
sums can be written as
\bea
&&
P^1_l(x,q) \eql \sum_{n\ {\rm even}}\ 
  x^n q^{{1 \over 2}n^2}
  \left[ \begin{array}{c} l \\ n \end{array} \right] \,,
\nonu
&&
P^2_{l+{1 \over 2}}(x,q) \eql \sum_{n\ {\rm odd}} \ 
  x^n q^{{1 \over 2}n^2}
  \left[ \begin{array}{c} l+1 \\ n \end{array} \right] \,,
\label{eqBCa}
\eea
whereas the $N$-fermion cut of the full chiral Hilbert space simply
reads
\be
{\rm ch}^{(N)}(q) \eql { q^{{1 \over 2} N^2} \over (q)_N} \,.
\label{eqBCc}
\ee
Clearly, both lead to the well-known Virasoro characters 
\bea
\chi_{(1,1)}(q) & \eql & \lim_{l\to\infty} P^1_l(x=1,q) \eql
  \sum_{N\ {\rm even}} \ {\rm ch}^{(N)}(q)\,, \nonu 
\chi_{(2,1)}(q) & \eql & \lim_{l\to\infty} P^2_{l+{1\over2}}(x=1,q) \eql
  \sum_{N\ {\rm odd}} \ {\rm ch}^{(N)}(q)\,. \label{eqBCd}
\eea

Upon contemplating expressions for `truncated partition sums'
and `$N$-particle cuts' one clearly wants to look for guidance in the 
extensive literature on CFT character formulas. In addition to the 
canonical `bosonic sum formulas' for the characters irreducible
highest weight modules of affine Lie algebras, of the Virasoro 
algebra, and and of $\cW$-algebras, there exist a variety 
of `fermionic sum formulas' and `bosonic product formulas'. The 
relations among the three types take the form of so-called 
Rogers-Ramanujan (RR) identities, which go back to the 19th century. 
In the process of proving some of the RR identities, several groups 
\cite{Schu,McM,ferchar2} introduced so-called $L$-finitizations of 
fermionic 
and bosonic character expressions and considered recursion relations 
satisfied by such finitized characters (in the affine Lie algebra case
a natural finitization of the character is provided by Demazure 
modules, see, e.g., \cite{FMO}).  In special cases, the
$L$-finitized characters agree with the `truncated partition sums' 
of section 2.1, and closed form expressions, both of the `fermionic sum' 
and of the `bosonic product' type are immediately available. We would 
like to stress that the existing literature on $L$-finitizations only 
covers some special cases of the 
`CFT statistics' program that we are pursuing here. 

The truncated partition sums for various level-1 WZW models are
special polynomials, whose structure is largely dictated by the
underlying Lie algebra structure. For the spinon formulation
of the $\fsu(n)_1$ WZW CFT, the truncated characters have been 
identified with the full partition sum of a so-called $\fsu(n)$ 
Haldane-Shastry spin chain on a finite number of sites.
In earlier papers \cite{BS1}, we provided explicit 
formulas for truncated characters and $N$-particle cuts for
$\fsu(n)_1$ spinons. Further character formulas for the 
$\fsu(n)_1$ and $\fso(n)_1$ WZW models will be presented in 
sections 5 and 6.

\newsection{Generalized fermions in minimal models}

A first generalization of the Majorana fermion is encountered in the unitary 
minimal model $\cM^m$ of central charge $\ccft(m)=1-{6 \over m(m+1)}$ with 
$m=3,4,\ldots$. We choose $\Phi_{(2,1)}$, of conformal dimension 
$h_{(2,1)}={m+3 \over 4m}$ as the fundamental quasi-particle and study the 
Virasoro sectors labeled as $(r,s)= (1,1), (2,1), \ldots, (m-1,1)$, that are 
generated by the repeated action of the modes of $\Phi_{(2,1)}(z)$ on the 
vacuum (cf.\ \cite{BPZ})
In the example $m=4$ we obtain the recursion relations
\be
\left( \begin{array}{c} P^1_{l+1} \\  P^2_{l+\half} \\ P^3_{l+1} \end{array} 
\right)
=
\left( \begin{array}{ccc} 
1	&  x  q^{l+{9 \over 16}} & x^2 q^{2l+\half} \\
x q^{l+{7 \over 16}} & 1 & x q^{l-{1 \over 16}} \\
0 & xq^{l+{1 \over 16}} & 1 
\end{array} \right)
\left( \begin{array}{c} P^1_{l} \\  P^2_{l-\half} \\ P^3_{l} 
\end{array} \right) \,, \label{blah}
\ee
where $P^r_l(x,q) = P^{(r,1)}_l(x,q)$.
For $q=1$, the associated eigenvalue $\lambda(x)$ satisfies
\be
(1-\lambda)^3+x^4-2x^2(1-\lambda) \eql 0\,,
\ee
and we read off that  (i) $\lambda(x)=1+\sqrt{2}x+{\cal O}(x^2)$ for small
$x$ and (ii) $\lambda(x) \sim  x^{4 \over 3}$ for $x$ large. As 
explained in section 2, the result (i)
has its origin in the non-abelian statistics of the field $\Phi_{(2,1)}$:
the degeneracy factor $\alpha=\sqrt{2}$ can be understood, as explained 
in section 2.1, as the largest
eigenvalue of the fusion matrix $N$ of the field $\Phi_{(2,1)}$ 
\be
N \eql  \left( \begin{array}{ccc}
0 & 1 & 0 \\
1 & 0 & 1 \\
0 & 1 & 0 
\end{array} \right) \ .
\ee
The exponent ${4 \over 3}$ in (ii) gives the maximal occupation of the
effective one-particle levels for $\Phi_{(2,1)}$ and sets the maximum
$n^{\rm max}$ of the associated generalization of the Fermi-Dirac 
distribution.

For general $m$, the recursion matrix (for $q=1$) has the form
\be \left( 
\begin{array}{ccccccc} 
1  &    x    &   x^2   &      &         &         &    \\
x  &    1    &    x    &  0   &         &         &   \\
0  &    x    &    1    &  x   &   x^2   &         &   \\
   &  x^2    &    x    &   1  &   x     &   0     &    \\
   &         &    0    &   x  &   1     &   x     &   \\
   &         &         & \ddots  & \ddots & \ddots & \ddots 
\end{array} 
\right) \,,
\label{rec-m}
\ee
and by inspecting the characteristic equation one obtains
\be
\alpha_m \eql 2 \cos ({ \pi \over m}) \,,
\qquad
n^{\rm max}_m \eql 2 \left( {m-2 \over m-1} \right) \,.\label{eqCe}
\ee
Note that in the limit $m\to\infty$,  $\alpha=2$ and $n^{\rm max}=2$, 
in agreement with the values for the $\fsu(2)_1$ spinons. In this sense, 
the $\Phi_{(2,1)}$ quasi-particles interpolate between the Majorana 
fermion and the $\fsu(2)_1$ spinons.

We remark that the recursion relations (\ref{rec-m}) are
mathematically identical to an iteration of the recursion relations
used by Andrews, Baxter and Forrester (ABF) in their analysis of local
height probabilities in specific RSOS models
\cite{ABF}. Interestingly, the two situations are dual in the sense
that, where the ABF recursions are in terms of the system size $m$,
the recursion relations (\ref{blah}), (\ref{rec-m}) are in terms of a
momentum variable $l$.

The systematics that led us to the recursion matrix (\ref{rec-m}) are
reminiscent of the structure underlying the `fermionic sum'
expressions for the characters of the minimal model $\cM^m$
\cite{ferchar1}.  Indeed, the truncated characters, which solve the
recursion relations set by (\ref{rec-m}), can be represented as
`fermionic sums'. We refer to \cite{GS} for an alternative point of
view on the results presented in this section.

\newsection{$\sZZ_k$ parafermions}

We move on to an alternative generalization 
of the Majorana fermion, namely the $\ZZ_k$ parafermion. 
This generalization arises if we replace the Virasoro algebra by a
$\cW_k$ algebra and focus on the simplest unitary minimal 
model with that symmetry. This model, at central charge 
$c_k={2(k-1) \over k+2}$, contains a primary field of conformal dimension 
$h={k-1 \over k}$ and this provides a natural generalization of  the 
Majorana fermion at $k=2$.

One possible quasi-particle formulation uses as fundamental quanta the 
modes of a set of $k-1$ parafermion fields $\psi^{(i)}$, $i=1,\ldots,k-1$, 
of conformal dimension $h^{(i)} = {i(k-i)\over k}$ and 
$\ZZ_k$ charge equal to $i$.  On the basis of the 
Lepowski-Primc character formulas \cite{LP}
(cf.\ (\ref{eqDia})), it has been argued  
that these quanta satisfy Haldane statistics with  matrix $G$  equal to 
twice the 
inverse of the Cartan matrix of the Lie algebra $A_{k-1}$ \cite{Ga}.
Indeed, we have argued in section 2.2 that both lead to the same central 
charge $\ccft$.  In this case the explicit solution of (\ref{eqBq})
is given by
\be 
x_a \eql 
\left( {\sin\left( { \pi\over k+2 } \right) \over 
  \sin\left( { \pi (a+1)\over k+2 } \right)} \right)^2 \,,
 \qquad a=1,\ldots,k-1\,,
\label{eqDaa}
\ee
and indeed
\be
\sum_{a=1}^{k-1}\ L(x_a) 
\eql \left( {\pi^2\over6} \right) \left( {2 (k-2)\over k+2 } \right) \eql 
\left( {\pi^2 \over 6} \right) \ccft \,,
\ee
by a well-known identity for dilogarithms (see, e.g., \cite{Lew,RS}).

The parafermion fields $\psi^{(i)}$ satisfy abelian braiding statistics 
and it is therefore possible to derive a set of generalized commutation 
relations for the modes $\psi^{(i)}_s$ \cite{ZF}. By exploiting these 
algebraic relations one may eliminate all fields with $i>1$ and generate 
the chiral spectrum by using the modes $\psi_{-s} \equiv
\psi^{(1)}_{-s}$ only.
For definiteness we first discuss the case $k=3$ where
the systematics of the construction are as follows.  Denoting
the modes of the single parafermion field by $\psi_{-s}$ we consider the 
combinations 
\be
\phi^{(1)}_{-s} \eql \psi_{-s} \,, \qquad
\phi^{(2)}_{-s} \eql \psi_{-s} \psi_{-s-{2 \over 3}} \,.\label{eqDa}
\ee
The states that we allow are of the form
\be
\phi^{(i_N)}_{-s_N} \ldots \phi^{(i_2)}_{-s_2} \phi^{(i_1)}_{-s_1}\,\vac\,,
\label{eqDb}
\ee
with minimal spacing specified as
\be
\begin{array}{cccc}
\stru {\rm if} & i_{l+1}=1,\ i_l=1 & {\rm then} & 
s_{l+1}-s_l ~\in~ \ZZ_{\geq0} + {1\over 3}  \\
\stru {\rm if} & i_{l+1}=2,\ i_l=1 & {\rm then} & 
s_{l+1}-s_l ~\in~ \ZZ_{\geq0} + {2 \over 3}  \\
\stru {\rm if} & i_{l+1}=1,\ i_l=2 & {\rm then} & 
s_{l+1}-s_l ~\in~ \ZZ_{\geq0} + {4 \over 3}  \\
\stru {\rm if} & i_{l+1}=2,\ i_l=2 & {\rm then} & 
s_{l+1}-s_l ~\in~ \ZZ_{\geq0} + {2 \over 3}  
\end{array} \label{eqDc}
\ee
and where $s_1\in\ZZ_{\geq0}+{2\over3}$ if $i_1=1$ and $s_1\in\ZZ_{\geq0}$
if $i_1=2$.

The truncated characters $X_l^{(i)}$ are defined by the restriction 
that the highest occupied mode is of type $\phi^{(i)}_{-s}$
with $l-s\in\ZZ_{\geq0}$. The above rules lead to the 
recursion relations
\bea 
X^{(1)}_{l+1}  -  X^{(1)}_{l} & \eql & 
xq^{{l+1}}\left( X_{l+{2\over3}}^{(1)}+X^{(2)}_{l-{1\over3}}\right)  \,,
\nonu
X^{(2)}_{l+1} - X^{(2)}_l & \eql & 
x^2 q^{{2l+ {8\over 3}}}\left(X_{l+{1\over3}}^{(1)}+X^{(2)}_{l+{1\over3}}
\right)  \,,
\label{eqDd}
\eea
with starting point $X^{(1)}_l = X^{(2)}_l = 0$, for $l\leq-1$, and
$X^{(1)}_{-{2\over3}}=0$, $X^{(2)}_{-{2\over3}}=1$.
Defining $Y_l = X^{(1)}_l + X^{(2)}_{l}$, we obtain
\be
Y_{l+1} \eql xq^{l+1} Y_{l+{2\over3}} + x^2 q^{2l +{8\over 3}} Y_{l+{1\over3}}
+ (1-x^3 q^{3l+3}) Y_{l}\,,\label{eqDe}
\ee
which, for $q=1$ and with $Y_l \sim \la^{l}$, leads to 
\be
\mu^3 - x \mu^2 - x^2 \mu - (1-x^3) \eql 0 \,,
\label{eqDf}
\ee
where $\mu = \la^{1\over3}$.
The associated statistics are abelian ($\alpha_{k=3}=1$), 
allow a maximal occupation of $n_{k=3}^{\rm max}=3$ and lead to 
the correct central charge of $c_{k=3}={4 \over 5}$. 
In \cite{Ga}, the eqn.\ (\ref{eqDf}) for $\lambda$ was recovered in
an approach which starts from the TBA equations for the
two $\ZZ_3$ parafermions $\psi^{(1)}$ and $\psi^{(2)}$,
leading to one particle partition functions $\lambda_1$ and
$\lambda_2$, and then performing the reduction to a single 
quasi-particle partition function by
$\lambda = \lambda_1 \lambda_2^2$ where, moreover, we have to take
$z_2=z_1^2 = x^2$.

Note that, for $x=1$ and $q=1$, the recursion relation (\ref{eqDe}) is
solved by $Y_{l/3} = F_l$, where $F_l$ is the $l$-th Fibonacci number
given explicitly, e.g., by Lucas' expression
\be
F_l \eql \sum_{k=0}^{[{l/2}]} \ \bin{l-k}{k} \,. \label{eqDfa}
\ee
For general $q$, the solution is a deformation of $F_l$.

Using the generalized commutation relations for the parafermions
$\psi^{(i)}$ one may write down yet another basis of the irreducible 
parafermion 
module in terms of the $\phi^{(i)}$.  One possible choice for such a 
basis is
\be 
\phi^{(1)}_{-s_{N_1}} \ldots \phi^{(1)}_{-s_{1}} \phi^{(2)}_{-t_{N_2}}
\ldots \phi^{(2)}_{-t_1} |0\rangle\,, \label{eqDfb}
\ee
where the sequences $\{s_i\}$ and $\{t_j\}$ each satisfy the 
conditions (\ref{eqDc}), $t_1\in\ZZ_{\geq0}$, and $s_1\in
\ZZ_{\geq0}+{2\over3} (N_2+1)$.
The basis (\ref{eqDfb}) immediately leads to the following 
parafermion character 
\be
{\rm ch}(x,q) \eql \sum_{n_1,n_2\geq0} \ x^{n_1+2n_2} \, 
  {q^{{1\over6}(n_1^2 + 4n_1n_2+
  4n_2^2 + 3n_1)} \over (q)_{n_1} (q^2)_{n_2}} \,, \label{eqDfc}
\ee
which can be shown to equal the Lepowski-Primc formula \cite{LP}
\be 
{\rm ch}(x,q) \eql \sum_{n_1,n_2\geq0} \ x^{n_1+2n_2} \, 
  {q^{{2\over3}(n_1^2 + n_1n_2+
  n_2^2)} \over (q)_{n_1} (q)_{n_2}} \,. \label{eqDfd}
\ee

For $\ZZ_k$ parafermions with general $k\geq 2$ we similarly 
combine the $\psi = \psi^{(1)}$ modes into combinations $\phi^{(i)}$,
$i=1,\ldots,k-1$, according to
\be
\phi^{(i)}_{-s} \eql \psi_{-s} \psi_{-s-{2\over k}}\ldots
\psi_{-s - {s(i-1)\over k}} \,. \label{eqDh}
\ee
The truncated partition sums $X_l^{(i)}$,
$i=1,\ldots,k-1$, $l\in\ZZ/k$, correspond to all states in
a quasi-particle basis built from the $\phi^{(i)}$ modes such that 
the highest occupied mode is of type $\phi^{(i)}_{-s}$ with
$s-l\in\ZZ_{\geq0}$.  We find the following recursion relation 
generalizing (\ref{eqDd})
\be
X^{(i)}_{l+1} - X^{(i)}_{l} \eql
x^i\, q^{i(l+1)+ {i(i-1)\over k}}\, \sum_{j=1}^{k-1} \
X^{(j)}_{l+1 - {2(k-i)\over k} + \delta_{i+j\leq k-1} } \,, \label{eqDi}
\ee
with starting point $X^{(i)}_l=0$, for $l\leq -1$, and 
$X^{(i)}_{-1+{1\over k}}= \delta_{i,k-1}$.
As for $k=3$ the recursion relation for the $X_l^{(i)}$ can be cast into
a single recurrence relation for $Y_l = \sum_i X^{(i)}_l$.  Note that,
in contrast to the case $k=3$, the different `sectors' $l\ \mod\ {1\over k}$
are no longer in 1--1 correspondence to the sectors with fixed $\ZZ_k$ 
charge.  We have checked numerically that the solution for 
\be 
Z_l \eql \sum_{j=0}^{k-1} Y_{l-{j\over k}}\,,\label{eqDib}
\ee 
in the limit $l\to\infty$ indeed approaches the Lepowski-Primc character
\be 
{\rm ch}(x,q) \eql \sum_{n_1,\ldots,n_{k-1}\geq0} \ \left( \prod 
  x^{in_i} \right) \,
  {q^{ {1\over 2} \sum n_i G_{ij} n_j } \over 
\prod_{i=1}^{k-1} (q)_{n_i} } \,, 
 \label{eqDia}
\ee
where $G_{ij}$ is twice the inverse Cartan matrix of $A_{k-1}$.

The equation for $Y_l$ leads to an equation for $\mu=\la^{1\over k}$ 
defined through $Y_l \sim \la^l$ as in (\ref{eqDf}), e.g.,
\bea
k\eql4\qquad && \mu^4 - x(1+2x^2) \mu^2 - (1+x^2)(1-x^4) \eql 0 \,,\nonu
k\eql5\qquad && \mu^5- x^2 \mu^4- 2x^4 \mu^3 -x(1-2x^5) \mu^2 
    - x^3 (1-x^5) \mu - (1-x^5)^2 \eql 0 \,. \nonu
&& \label{eqDj}
\eea
We find, for arbitrary $k\geq2$,
\be
\alpha_k\eql 1 \,, \qquad n_k^{\rm max} \eql {k(k-1) \over 2} \,. 
\label{eqDg}
\ee
We have checked, for $k=4,5$, that the equations (\ref{eqDj}) agree
with the equation for $\la = \la_1 \la_2^2 \ldots \la_{k-1}^{k-1}$
starting from the TBA equation (\ref{eqBj}) for $k-1$ parafermions
$\psi^{(i)}$ with a statistical interaction matrix $G_{ij}$ given by
twice the inverse Cartan matrix of the Lie algebra $A_{k-1}$ and 
fugacities $z_i = x^i$, thus
confirming the conjecture of \cite{Ga}.

\newsection{$\sfsu(n)_1$ WZW model}

We start the discussion of exclusion statistics for conformal 
field theories based on WZW models with the case of $\fsu(n)_1$
(see also section 2.3 for $\fsu(2)_1$).

Let $\La_a$, $a=1,\ldots,n-1$, denote the fundamental weights of $\fsu(n)$.
We denote by $\chi_{\La_a}$ the (formal) character of the 
finite dimensional irreducible representation 
$L(\La_a)$ of $\fsu(n)$ with highest weight $\La_a$.  
Evaluating the character $\chi_{\La_a}$ at the identity gives the 
dimension of $L(\La_a)$ 
\be 
{\rm dim\,}L(\La_a) \eql \bin{n}{a} \,, \qquad a=1,\ldots,n-1\,.
\label{eqEAa}
\ee
The affine Lie algebra $\fsu(n)_1$ has $n$ integrable highest weight modules
corresponding to highest weights $\La_0, \La_1, \ldots, \La_{n-1}$ with
conformal dimensions
\be
h(\La_a) \eql { a(n-a)\over 2n} \,, \label{eqEAb}
\ee
while the central charge of $\fsu(n)_1$ is given by
\be 
\ccft \eql n-1 \,.\label{eqEAba}
\ee
In accordance with the conventions in \cite{BS1} let us take the
fundamental quasi-particle (spinon) to transform in the irrep
$L(\La_{n-1}) = \bar{{\bf n}}$ of $\fsu(n)$.  Let us define the
truncated characters $P^{(a)}_l(x,q)$, $a=0,\ldots,n-1$, as in
(\ref{eqBb}), where we have assigned the same fugacity $x_i=x$ to all
spinons in the $\bar{\bf n}$.  We will keep track of the $\fsu(n)$
weights as well so, strictly speaking, the $P^{(a)}_l(x,q)$ are
character valued polynomials in both $x$ and $q$.
The recursion relations for the $P^{(a)}_l(x,q)$ follow straightforwardly
from the spinon basis constructed in \cite{BS1}.  In terms of 
the character valued polynomials $X_l(x,q)$, $l\in\ZZ/n$, where
\be
X_l \eql P_l^{(a)} \,, \qquad {\rm for}\quad  nl \equiv a\, \mod \,1\,,
\label{eqEAbb}
\ee
they are given by
\be
X_l \eql \sum_{m=1}^n\ x^{n-m}q^{ (n-m)(l-{m\over2n}) }\ 
\left(\prod_{i=1}^{m-1} (1-x^nq^{nl-i})\right)\ 
\chi_{\La_m} \ X_{l-{m\over n}}\,,
\label{eqEAc}
\ee
with $\Lambda_n\equiv\Lambda_0$. The starting point for the recursion is
$X_l=0$ for $l<0$ and $X_0=1$.
Specializing the character $\chi_{\La_m}$ to the dimension of $L(\La_m)$
(cf.\ eqn.\ (\ref{eqEAa}))
and putting $q=1$, gives the following equation for $\la_{\rm av}(x)$, where
$X_l(x;1) \sim \la_{\rm tot}(x)^l = \la_{\rm av}(x)^{nl}$
\be 
\la_{\rm av}^n - \sum_{m=1}^{n} \left( \begin{array}{c} n \\ 
m\end{array} \right) 
(1-x^n)^{m-1} x^{n-m} \la_{\rm av}^{n-m} = 0\,, \label{eqEAd}
\ee
or, equivalently, 
\be
1 - ( x + (1-x^n) \la_{\rm av}^{-1} )^n \eql 0\,. \label{eqEAe}
\ee
The physical solution of (\ref{eqEAe}) is given by 
\be 
\la_{\rm av}(x) \eql { 1-x^n\over 1-x} \eql 1+ x + \ldots + x^{n-1}\,.
\label{eqEAf}
\ee
The large and small $x$ limits can be immediately read off and lead to
\be 
\al\eql 1 \,,\qquad n_{\rm av}^{\rm max} \eql n-1 \,. \label{eqEAg}
\ee
The statistics going with the distribution (\ref{eqEAf}) generalize the
fermionic statistics of section 2.3 in the sense that a state can contain 
at most $n_{\rm av}^{\rm max} = n-1$ excitations 
with the same quantum numbers.
These kind of statistics were proposed by Gentile as early as 1940 
\cite{Gen}.

Along the same lines one can show that a single quasi-particle species,
in the absence of the others, behaves as a $g$-on with $g=(n-1)/n$, i.e.,
we have (cf.\ section 2.2)
\be
n^{\rm max}_a \eql {1\over g} \eql {n\over n-1} \,.\label{eqEAga}
\ee

As a consistency check on the recursion (\ref{eqEAc}) we can compute
the central charge $\ccft$ through eqns.\ 
(\ref{eqBe}) and (\ref{eqBf}).  Indeed, we find
\bea
{\pi^2 \over 6} \ccft & \eql &  \int_0^1 \ {dx\over x} 
\log\la_{\rm tot}(x) \eql n \int_0^1 \ {dx\over x} 
\log\la_{\rm av}(x)\nonu
 & \eql & n \int_0^1 \ {dx\over x} \log(1-x^n) - 
  n \int_0^1 \ {dx\over x} \log(1-x) \nonu
 & \eql & n ( -{1\over n} +1 ) \ {\rm Li}_2(1) \eql {\pi^2 \over 6} (n-1) \,,
\label{eqEAh}
\eea
where ${\rm Li}_2(x)$ was defined in (\ref{eqBp}) and we have 
used the value ${\rm Li}_2(1) = {\pi^2 \over 6}$.  

By putting $x=1=q$ in (\ref{eqEAc}) we find 
\be 
X_{l}(1,1) \eql \left(\chi_{\La_1} \right)^{nl}\,, \qquad
nl\in\ZZ_{\geq0}\,.
\label{eqEAj}
\ee
Note that $X_{l}(1,1)$ is precisely the representation content of 
a one-dimensional spin chain of length $L=nl$ where the spins transform
in the irrep $L(\La_1) = {\bf n}$ of $\fsu(n)$.  
For general $(x,q)$ the solution $X_l(x,q)$ of (\ref{eqEAc}) will be a 
deformation of (\ref{eqEAj}).
One might wonder whether $X_l(1,q)$ can be identified with the 
partition function of such a spin chain.  This indeed turns out to be the
case, $X_{l}(1,q)$ corresponds precisely to the partition function 
of the $\fsu(n)$ Haldane-Shastry spin chain of length $L=nl$ \cite{Ha3,Sh} 
(this partition function was computed in, e.g., \cite{BS1}, eqn.\ (3.12)).

A natural deformation of (\ref{eqEAj}) is the
(dual) Milne polynomial $M_\la(x_a,q)$ defined in Appendix A, i.e.,
\be 
M_\la(x_a,q) \eql \sum_\mu \ \left( \prod_a x_a^{(\mu,\al_a^\vee)} \right)\, 
\wK_{\mu\la}(q) \chi_{\mu}\,, \label{eqEAk}
\ee
where $\wK_{\la\mu}(q)$ is the dual Kostka polynomial defined in Appendix A.
Indeed, we find the following solution
\be
X_{l}(x,q) \eql q^{ {1\over2} n(n-1)l^2 } M_{l \La_1} (x,q^{-1}) \,,
\qquad nl\in \ZZ_{\geq0}\,,
\label{eqEAl}
\ee
where $x_a = x^{n-a}$, $a=1,\ldots,n-1$.
An explicit formula for the dual Milne polynomial entering 
in (\ref{eqEAl}), at $x=1$, is (cf.\ \cite{Kir}, section 2.4.1)
\bea
M_{m\La_1}(1,q) & \eql & \sum_{m_1,\ldots,m_n\geq0\atop m_1+\ldots+m_n=m}\
\qbin{m}{m_1,\ldots,m_n} e^{m_1\eps_1+  \ldots +m_n\eps_n} \nonu
& \eql & (q)_m \sum_{m_1,\ldots,m_n\geq0\atop m_1+\ldots+m_n=m}\
{1\over (q)_{m_1} \ldots (q)_{m_n}}\, e^{m_1\eps_1+\ldots+m_n\eps_n}
\,,\label{eqEAm}
\eea
where the $\eps_i$, $i=1,\ldots,n$, are the weights of the $n$-dimensional
irreducible representation $L(\La_1)$ of $\fsu(n)$.  Clearly, (\ref{eqEAm})
has an interpretation as ($(q)_m$ times) the $m$-particle cut of the
Fock space character of a set of $n$ quasi-particles transforming in 
the irreducible 
representation $L(\La_1)$ of $\fsu(n)$.  It is possible to give a formula 
analogous to (\ref{eqEAm}) for the Milne polynomial $M_\la(1,q)$ 
corresponding
to a general weight $\la$ by introducing a set of quasi-particles
for each fundamental irreducible representation $L(\La_a)$ of $\fsu(n)$
(see, e.g., section 5.1 for $\fsu(3)$).  In terms of Milne polynomials, 
the $N$-particle decomposition of the affine $\fsu(n)_1$ characters 
is given by (cf.\ \cite{NYb} for $x=1$)
\be
{\rm ch}(x,q) \eql \sum_{N\geq0}\ x^N\, {\rm ch}^{(N)}(q) \eql
\sum_{ \mu = \sum m^a \La_a} \
\left( \prod_a x^{(n-a)m^a} \right) \ 
{q^{ {1\over2} |\mu|^2}  \over \prod_{a=1}^n (q)_{m^a} } 
M_\mu (q) \,. \label{eqEAn}
\ee
The $N$-particle cut ${\rm ch}^{(N)}(q)$ contributes to the affine
character in the sector $L(\La_{-N\, \mod\, n})$.
Equation (\ref{eqEAn}) 
emphasizes the fact that the quasi-particles transforming 
in representations $L(\La_a)$ should be viewed as composites
of $n-a$ elementary `spinons' transforming in the representation 
$L(\La_{n-1})$.
Indeed, the $\fsu(n)_1$ modules can be built up using the $L(\La_{n-1})$ 
spinons only (cf.\ section 4).  
An explicit formula for ${\rm ch}(x,q)$, using only the $L(\La_{n-1})$ 
spinons, was given in \cite{BS1}
\be 
{\rm ch}(x,q) \eql  \sum_{\mu=-\sum m_i\eps_i } \ \left( \prod_i x^{m_i} 
\right) \ 
\sum_{p\geq0} (-1)^p {q^{ {1\over2}p(p-1)} \over (q)_p }
{ q^{ {1\over2} |\mu|^2}\over \prod_i (q)_{m_i-p} }\,
e^{ \mu } \,. \label{eqEAo}
\ee

Finally, we recall that both the $\fsu(n)_1$ modules and their  
truncations (whose characters are given by $X_l(1,q)$) admit an
action of the Yangian $Y(\fsu(n))$.  This action finds it origin
in the Haldane-Shastry spin chain \cite{HHTBP,BGHP}.  The decomposition of
both the affine characters and their truncations $X_l(1,q)$ under 
the action of $Y(\fsu(n))$ were discussed in \cite{BS1,KKNa}.

\subsection{$\sfsu(3)_1$}

In the previous section we remarked that an explicit formula for the Milne
polynomial $M_\la(q)$ entering the $N$-particle characters (\ref{eqEAn})
can be given by introducing quasi-particles
for each fundamental irreducible representation $L(\La_a)$ of $\fsu(n)$.
Here we make this more explicit for $\fsu(3)$ and remark on the origins
of these formulas.

For $\fsu(3)$ we introduce two sets of three quasi-particles, transforming in 
the $L(\La_1) = {\bf 3}$ and the $L(\La_2) = {\bf 3}^*$, respectively.
The character of the subspace of the total Fock space containing $m^a$ 
particles of type $a$ ($a=1,2$) is given by
\be
\cM_{(m^1,m^2)}(q) \eql
\sum_{ \sum_i m_i^a = m^a }
\ {1\over \prod_{a=1}^2 \prod_{i=1}^3 (q)_{m_i^a} }
e^{ \sum_i (m_i^1 - m_i^2)\eps_i }\,.\label{eqEBa}
\ee
In terms of the quasi-particle Fock space characters $\cM_{(m^1,m^2)}(q)$
the most general $\fsu(3)$ Milne polynomial at $x_a=1$ can be expressed as 
\be 
M_{m^1\La_1+m^2\La_2}(1,q) \eql (q)_{m^1}(q)_{m^2}\sum_{p\geq0} (-1)^p
{q^{ {1\over2} p(p-1)} \over (q)_p}  \cM_{(m^1-p,m^2-p)}(q) \,.
\label{eqEBb}
\ee
The equality of (\ref{eqEAn}) and (\ref{eqEAo}) for $\fsu(3)$ (using
eqn.\ (\ref{eqEBb})) was demonstrated in \cite{BS1}.

A similar formula holds for the $\fsu(n)$ Milne polynomial
$M_{m^1\La_1+m^n\La_n}(1,q)$.  The quasi-particle expression for general
$\fsu(n)$ weight $\la$ is considerably more complicated.

In subsequent sections we will see other examples of formulas of the 
type (\ref{eqEAm}) and (\ref{eqEBb}).  Let us briefly remark on the
origins of these formulas.
Consider a finite dimensional simple (complex) 
Lie algebra $\bfg$ of rank $\ell$
and with fundamental weights $\La_a$, $a=1,\ldots,\ell$.  Introduce a set
of coordinates $x^{(a)}_i$, $i=1,\ldots,{\rm dim\,}L(\La_a)$,
for each fundamental irreducible representation $L(\La_a)$.  The Lie
algebra $\bfg$ acts on the polynomial ring $\CC[x^{(a)}_i]$ by
linear differential operators, preserving the subspaces 
$\CC[x^{(a)}_i]_{ \{ m^a\} }$ consisting of homogeneous polynomials of
order $m^a$ in the $x^{(a)}_i$.  In general, though, the algebra does
not act irreducibly on $\CC[x^{(a)}_i]_{\{m^a\} }$, but preserves an ideal
$I$ generated by homogeneous relations.  (For example, in the case
of $\fsu(3)$ above the ideal $I$ is generated by $\sum_i x_i^{(1)} 
x_i^{(2)}$.)  In fact, $\CC[x^{(a)}_i]/I$
is the coordinate ring of a flag manifold associated to $\bfg$. 
The action of $\bfg$ on $\CC[x^{(a)}_i]_{\{m^a\}} / 
(\CC[x^{(a)}_i]_{ \{m^a\} }\cap I)$
is irreducible and the irreducible representation is isomorphic to
$L(\sum m^a \La_a)$.  The character of $L(\sum m^a \La_a)$ follows 
easily once we have a free resolution of $\CC[x^{(a)}_i]/I$ by
applying the Euler-Poincar\'e principle.  Equations like 
(\ref{eqEAm}) and (\ref{eqEBb}) can be interpreted as `affinizations' of
the above constructions where, instead of the
polynomial ring $\CC[x_i^{(a)}]$, we
have a Fock space of quasi-particles $\phi^{(a)}_i(z)$.  
We refer to \cite{Bou} for more details, see also \cite{FS} for 
closely related ideas.

\subsection{$\fsu(2)_k$, $k\geq1$}

The spinon basis for $\fsu(2)_k$, $k\geq1$, has been worked out
in \cite{BLS2}.  Here also the $N$-spinon cuts of the 
affine characters were found.  
These were subsequently proved in \cite{NYa,ANOT}.
The $\fsu(2)_k$ recursion relations were written down in \cite{FrS}.
For completeness we briefly review these results.  We refer to
\cite{BLS2,FrS} for more details.

The affine Lie algebra $\fsu(2)_k$ has $k+1$ integrable highest weight 
modules, with highest weights $(k-i)\La_0 + i \La_1$, $i=0,\ldots,k$,
and conformal dimension
\be 
h(i) ~\equiv~ h((k-i)\La_0+i\La_1) 
  \eql { i(i+2) \over 4(k+2) } \,. \label{eqECa}
\ee
The central charge of $\fsu(2)_k$ is given by
\be
\ccft \eql {3k\over k+2} \,. \label{eqECb}
\ee
As our fundamental quasi-particle we again take the spinon transforming
in the irrep $L(\La_1) = {\bf 2}$ of $\fsu(2)$ and we denote by 
$P_l^{(i)}(x,q)$, $i=0,\ldots,k$, $l\in\ZZ/2$, the truncated spinon 
character in the sector $L((k-i)\La_0 + i \La_1)$.
If we denote the 
character of the $\fsu(2)$ irrep $L(m\La_1) = {\bf m+1}$ by $\chi_m$
(note that $\chi_m = \chi_{{\bf m+1}}$ in the notation of section 2.3)
the recursion relations take the following form 
\bea
P_{l+{1\over2}}^{(i)} & \eql & x^i q^{ il + h(i)} \chi_i P_l^{(0)} +
  (1-x^2q^{2l}) \sum_{j=1}^i \ x^{i-j} q^{(i-j)l + h(i)-h(j)} 
  \chi_{i-j} P^{(j)}_l \,,\nonu
P_{l+1}^{(i)} & \eql & (1-x^2 q^{2l+1}) \sum_{j=i}^{k-1}\
   x^{j-i} q^{(j-i)(l+1) + h(i)-h(j)} \chi_{j-i} P^{(j)}_{l+{1\over2}} \nonu
 &&  \  + x^{k-i} q^{(k-i)(l+1) + h(i)-h(k) } \chi_{k-i} 
  P^{(k)}_{l+{1\over2}}\,,
\label{eqECc}
\eea
where $l\in\ZZ_{\geq0}$.  Alternatively, one may write the 
equations (\ref{eqECc}) in matrix form (cf.\ (\ref{eqBBb})) \cite{FrS}
\be 
\vec{P}_{l+1}(x,q)   \eql  {\cal R}_l(x,q) \vec{P}_{l}(x,q)\,,\label{eqECd}
\ee
from which it is clear that the grand partition function $\la_{\rm tot}(x)$
can be obtained as the largest eigenvalue of the matrix ${\cal R}_l(x,1)$.
Explicitly, after specializing the characters to the dimensions,
one finds for $\mu=\lambda_{\rm tot}^{1 \over 2}$
(see also \cite{FrS} for $k=2$)
\bea
k=2 && \quad (\mu-1)^2 - x^2 (\mu+1) \eql 0 \,,\nonu
k=3 && \quad (\mu-1)^2 -x (\mu-1) - x^3(\mu+1) -x^2 \eql 0 \,, \nonu
k=4 && \quad (\mu-1)^3 -x^2 (\mu-1)(\mu+2) - x^4 (\mu+1)^2 \eql 0 \,, \nonu
{\rm etc.} &&
\eea
The asymptotics of $\la_{\rm tot}(x)$ yield
\be
\al \eql 2 \cos \left( {\pi\over k+2} \right)\,,
\qquad n^{\rm max}_{\rm tot} \eql 2k\,,\label{eqECe}
\ee
signaling the presence of non-abelian exclusion statistics for $k\geq2$.

Clearly, the recursion (\ref{eqECc}) for $x=1=q$ is solved by 
\bea
P_{l+{1\over2}}^{(i)}(1,1) & \eql & \chi_{i} \left( \chi_k \right)^{2l}
  \,,\nonu
P_l^{(i)}(1,1) & \eql & \chi_{k-i} \left( \chi_k \right)^{2l-1} \,,\qquad
l\in\ZZ_{\geq0}\,.
\label{eqECf}
\eea
The solution for general $q$, $x=1$, and $l\in\ZZ_{\geq0}$, 
turns out to be given by
\bea
P_{l+{1\over2}}^{(i)} & \eql & 
q^{h(i) - kl(l-1) - il} Q'_{(k^{2l}\, i)}(q) \,, \nonu
P_l^{(i)} & \eql & 
q^{h(i)-kl(l-1)+i(l-1)} Q'_{(k^{2l-1}\, k-i)}(q)\,, 
\label{eqECg}
\eea
where we use the standard notation $(1^{m_1}\, 2^{m_2} \ldots)$ for 
partitions, and 
where $Q'_\la(q)$ is the Milne polynomial
defined in Appendix A, eqn.\ (\ref{eqZk}).  Namely,
\be
Q'_\la(q) \eql \sum_{\mu=(\mu_1,\mu_2)} \  
K_{\mu\la}(q)\, \chi_\mu \,.\label{eqECh}
\ee
Here the sum is over all partitions $\mu$ with at most two parts (in order
for the $\fsu(2)$ character $\chi_\mu$ to be nonvanishing), and
$K_{\la\mu}(q)$ is the Kostka polynomial.  Explicit expressions for the
Milne polynomials $Q'_{(k^{l})}(q)$ can be found in \cite{Kir},
Theorem 14.  That the $l\to\infty$ limit of the expressions (\ref{eqECg})
produces the $\fsu(2)_k$ affine characters has been established in
\cite{Kir,NY} for $i=0$.

\newsection{$\sfso(n)_1$ WZW model: free fermion CFT}

It is well-known that the integrable highest weight modules of affine
$\fso(n)$ at level-1 can be realized in terms of $n$ free Majorana
fermions transforming in the vector representation of $\fso(n)$.
In this section we will show that,
alternatively, one can realize these modules in terms of
quasi-particles transforming in the spinor representation(s) of
$\fso(n)$ -- these quasi-particles are referred to as spinons.  In a sense
they are more fundamental than the fermions since the latter can be 
expressed in terms of composites of spinons.  It will
be necessary to discuss the $n$ even or odd case separately.  First we
discuss the $n$ odd case, which is a generalization of the known
$\fso(3)_1 \cong \fsu(2)_2$ result, then we discuss $n$ even.  Several low
rank cases not covered by this analysis, nevertheless
interesting  in their own right, will be discussed separately.

\subsection{$\sfso(2n+1)_1$, $n\geq2$}

Let $\La_a\,,a=1,\ldots,n$, denote the fundamental weights of $\fso(2n+1)$.
The dimension of the finite dimensional irreducible representation 
$L(\La_a)$ of $\fso(2n+1)$ with highest weight $\La_a$ is given by
\be
{\rm dim\ } L(\La_a) \eql \left\{
\begin{array}{cl} 
\bin{2n+1}{a}  & {\rm for\ } a=1,\ldots,n-1\,,\\  & \\
2^{n} & {\rm for\ } a=n \,. 
\end{array} \right. \label{eqFAa}
\ee
The affine Lie algebra $\fso(2n+1)_1$ has three integrable highest
weight modules corresponding to highest weights $\La = \La_0, \La_1$ and
$\La_n$, referred to as the singlet ($\id$), vector ($v$) and spinor
($s$), respectively.  Here, and in the rest of this section, we take
$n\geq2$.  The cases $n=0,1$ will be treated separately.  
The conformal dimensions are given by
\be
h(\La_0) \eql 0\,,\qquad h(\La_1) \eql {1\over2}\,,\qquad
h(\La_n) \eql {2n+1\over16}\,, \label{eqFAb}
\ee
the central charge of $\fso(2n+1)_1$ is
\be
\ccft \eql {2n+1\over2} \,,\label{eqFAc}
\ee
and the fusion rules are given by
\be
s\times v \eql s\,,\qquad s\times s \eql 1+v\,,\qquad v\times v \eql 1\,.
\label{eqFAd}
\ee

Consider the $\fso(2n+1)_1$ module spanned by the modes
$\phi_{-s}^{(i)}$ of the $2^n$ spinon operators and let
$P_l^{(a)}(x,q)$ ($a=1,v,s$) be the truncated partition 
function
(\ref{eqBb}), where we have assigned the same fugacity $x_a=x$ to all
spinons.  
The recursion relations look most elegant in terms of the character 
valued polynomials $X_l$ and $Y_l$ ($l\in\ZZ/2$)
\bea
X_l & \eql & \left\{  \begin{array}{cl} 
P_l^{(\id)} & {\rm for\ } l \ {\rm integer} \,,\\
P_l^{(v)} &  {\rm for\ } l \ {\rm half\ odd\ integer} \,, \end{array} \right. 
\nonu
Y_l & \eql & \quad P_l^{(s)} \quad {\rm for\ all\ } l\,. \label{eqFAe}
\eea
We obtain the following recursion relations
\bea
Y_{l+{1\over2}} & \eql &  (1-x^2 q^{2l})Y_l +
x q^{ {2n+1\over 16} +l } \chi_{\La_n} X_l\,, \nonu
X_{l+{1\over2}} & \eql &  x^2 q^{2l+{1\over2}}  \chi_{\La_1} X_l +
(1-x^2 q^{2l})  \chi_{\La_0} X_{l-{1\over2}} \nonu
& & + x^2 \sum_{i=2}^{n-1} \left(  q^{2l + 1 - {i\over 2}} 
\left( \prod_{k=0}^{i-2} (1-x^2 q^{2l-k} ) \right)  \left(
\sum_{j=0}^{[i/2]} \chi_{\La_{i-2j}} \right)
 \right) X_{l - {i-1\over2}} \nonu
& & + x q^{-{2n+1\over 16} + l + {1\over2}}   \prod_{k=0}^{n-2} 
(1-x^2 q^{2l-k} ) \chi_{\La_n} Y_{l-{n-2\over2}} \nonu
& & - x^2 \sum_{i=1}^n \, q^{2l-{n-3+i\over2}}  \left(
\prod_{k=0}^{n-3+i} (1-x^2 q^{2l-k} ) \right) \left( 
\sum_{j=0}^{[(n-i)/2]} \chi_{\La_{n-i-2j}} \right) 
X_{l - {n+i-2\over2}} \,.\nonu
&&\label{eqFAf}
\eea
After substituting $X_l \sim \la^l_{\rm tot}$, 
$Y_l\sim \mu \la_{\rm tot}^{l-{1\over2}}$ 
in the recursion relation (\ref{eqFAf}), putting $q=1$ and specializing the 
characters to the dimension, we find
that the equation for the grand partition function $\la_{\rm tot}(x)$ 
possesses an (unphysical) root $\la_{\rm tot}^{1\over2}=-(1-x^2)$ 
for each $n$.  
After dividing
out this root, the equation can most succinctly be written as 
\be
(1-\ze)^2 \eql x^2 (1+\ze)^{2n-1}\,,\label{eqFAg}
\ee
where
\be
\ze \eql (1-x^2) \la_{\rm tot}^{-{1\over2}}\,.\label{eqFAh}
\ee
It follows immediately that 
\be 
\al \eql \sqrt{2}\,,\qquad
n^{\rm max}_{\rm tot} \eql 4\,.\label{eqFAi}
\ee
Note that the factor $\sqrt2$ arises indeed as the largest eigenvalue 
of the fusion matrix corresponding to (\ref{eqFAd}).

As a check on the recursion relations (\ref{eqFAf}) 
we can compute the resulting 
central charge through eqns.\ (\ref{eqBe}) and (\ref{eqBf}).
\bea
{\pi^2\over6} \ccft & \eql & \int_0^1 \ {dx\over x} \log\la \eql
  2 \int_0^1\ {dx\over x} \log(1-x^2) - 2 \int_0^1 \ {dx\over x} \log\ze \nonu
& \eql & 2 \int_0^1\ {dx\over x} \log(1-x^2) -
  2 \int_0^1\ {d\ze\over 1-\ze} \log\ze - 
  (2n-1) \int_0^1\ {d\ze\over 1+\ze} \log\ze \nonu
& \eql & \left( -1 + 2 + {\textstyle{1\over2}}(2n-1) \right) {\rm Li}_2(1) 
  \eql \left( {2n+1\over2} \right) {\pi^2\over6} \,,\label{eqFAj}
\eea
in accordance with (\ref{eqFAc}).

The recursion relations (\ref{eqFAf}) can be solved exactly.  
Observe, first of all, that for $x=1=q$ we have 
\be
X_l \eql \left( \chi_{\La_1} \right)^{2l} \,,\qquad
Y_l \eql \chi_{\La_n} \left( \chi_{\La_1} \right)^{2l-1} \,.\label{eqFAk}
\ee
For general $q$ the solution is a $q$-deformation of (\ref{eqFAk}).  
As for $\fsu(n)$ it turns out
that the correct $q$-deformation is again the Milne
polynomial $M_\la(x_a,q)$ defined in Appendix A, i.e.,
\bea
X_{l} & \eql &  q^{2l^2 }   M_{2l\La_1}(x,q^{-1}) \,, \nonu
Y_{l} & \eql & q^{ {2n+1\over 16} +   l(2l-1) } 
M_{(2l-1)\La_1+\La_n}(x,q^{-1})\,,\label{eqFAl}
\eea
where $x_a=x^2$ for $a=1,\ldots,n-1$ and $x_n=x$.
An explicit expression for the Milne polynomials entering in (\ref{eqFAl}),
at $x=1$,
can be given, namely (cf.\ (\ref{eqEAm}))
\bea
M_{l\La_1}(1,q) & \eql & (q)_l \sum_{p\geq0} (-1)^p {q^{{1\over2}p(p-1)}
\over (q)_p } \cM_{(l-2p,0)}(q) \,,\nonu
M_{l\La_1+\La_n}(1,q) & \eql & (q)_l (q)_1 \left( \sum_{p\geq0} (-1)^p 
{q^{{1\over2}p(p-1)} \over (q)_p} \cM_{(l-2p,1)}(q) \right.  \nonu  && -
\sum_{p\geq0} (-1)^p 
{q^{{1\over2}p(p+1)} \over (q)_p(q)_1} 
\cM_{(l-2p-1,0)}(q)\chi_{\La_n} \nonu && \left. +
\sum_{p\geq0} (-1)^p 
{q^{{1\over2}p(p+1)} \over (q)_p(q)_1} 
\cM_{(l-2p-2,0)}(q)\chi_{\La_n} \right) \,,
\label{eqFAm}
\eea
where $\cM_{(l,m)}(q)$ is the Fock space
character for quasi-particles transforming 
in the $(2n+1)$-dimensional vector- and $2^n$-dimensional spinor
representation of $\fso(2n+1)$, i.e.,
\be
\cM_{(l,m)}(q) \eql \sum_{l_1+\ldots+l_{2n+1} = l \atop m_1+\ldots+m_{2^n}=m}
\ {1\over (q)_{l_1} \ldots (q)_{l_{2n+1}} } {1\over
(q)_{m_1} \ldots (q)_{m_{2^n}} } e^{\sum_i l_i \la_i + \sum_\alpha m_\alpha
\tilde{\la}_\alpha} \,.
\label{eqFAn}
\ee
Here $\la_i$ and $\tilde{\la}_\alpha$ denote the weights of the
vector and spinor representation, respectively.
The second expression in (\ref{eqFAm}) can be simplified, but we have
left it in this form to elucidate its origin, i.e. it arises from the
resolution of the coordinate ring of a (partial) flag manifold associated 
to $\fso(2n+1)$ (see the remarks in section 5.1). 

As for $\fsu(n)$ (cf.\ eqn.\ (\ref{eqEAn})) it should be possible to 
write a formula for the $N$-particle cut of the 
affine $\fso(2n+1)_1$ characters in terms of Milne polynomials.  However,
due to the more complicated fusion rules of $\fso(2n+1)_1$ this formula is
somewhat more involved.  In fact, it is only known for $\fso(5)$ (see
\cite{Yam} and section 6.7).

\subsection{$\sfso(2n)_1$, $n\geq3$}

Let $\La_a\,,a=1,\ldots,n$, denote the fundamental weights of $\fso(2n)$.
The dimension of the finite dimensional irreducible representation 
$L(\La_a)$ of $\fso(2n)$ with highest weight $\La_a$ is given by
\be
{\rm dim\ } L(\La_a) \eql \left\{
\begin{array}{cl} 
\bin{2n}{a}  &  {\rm for\ } 1\leq a\leq n-2\,, \\ &\\
2^{n-1} & {\rm for\ } a=n-1,n \,. 
\end{array} \right. \label{eqFBa}
\ee
The affine Lie algebra $\fso(2n)_1$ has four integrable highest weight 
modules corresponding to highest weights $\La_0, \La_1, \La_{n-1}$ and 
$\La_n$.  Here, and in the rest of this section, we take $n\geq3$.  The
cases $n=1,2$ will be treated separately.  The conformal dimensions are
given by 
\be 
h(\La_0)\eql0\,,\qquad h(\La_1)\eql {1\over2}\,,\qquad
h(\La_{n-1}) \eql h(\La_{n}) \eql {n\over8}\,,
\label{eqFBb}
\ee
and the central charge is
\be 
\ccft\eql n\,.\label{eqFBc}
\ee

For convenience, let $s=L(\La_{n-1})$ denote the spinor,
$c=L(\La_n)$ the conjugate spinor and $v=L(\La_1)$ the vector 
representation of $\fso({2n})$.  As our fundamental quasi-particles 
we will now take both the $s$ and $c$ spinors.\footnote{Note that,
even though $\fsu(4)_1 \cong \fso(6)_1$, our description of $\fso(6)_1$
differs from the one for $\fsu(4)_1$ discussed in section 5.}
The fusion rules 
for $\fso(2n)_1$ depend on whether $n$ is even or odd.  For $n=2p$ the
relevant fusion rules are  
\be
s \times s \eql \id\,,\quad
c \times c \eql \id\,,\quad
s \times c \eql  v\,,\quad
s \times v \eql  c \,,\quad
c \times v \eql  s \,,
\label{eqFBd}
\ee
while for $n=2p+1$ we have 
\be
s \times s \eql v\,,\quad
c \times c \eql v\,,\quad
s \times c \eql \id \,\quad
s \times v \eql c\,,\quad
c \times v \eql  s \,.
\label{eqFBe}
\ee
The recursion relations are most elegantly written in terms of 
character valued polynomials $X_l, Y_l$ and $Z_l$ ($l\in\ZZ/2$)
defined by
\bea 
X_l & \eql & \left\{  \begin{array}{cl} 
P_l^{(\id)} & {\rm for\ } l \ {\rm integer} \,,\\
P_l^{(v)} &  {\rm for\ } l \ {\rm half\ odd\ integer} \,, \end{array} \right. 
\nonu
Y_l & \eql & \left\{  \begin{array}{cl} 
P_l^{(s)}  & {\rm for\ } l \ {\rm integer} \,,\\
P_l^{(c)} &  {\rm for\ } l \ {\rm half\ odd\ integer} \,, \end{array} \right. 
\nonu
Z_l & \eql & \left\{  \begin{array}{cl} 
P_l^{(c)}  & {\rm for\ } l \ {\rm integer} \,,\\
P_l^{(s)} &  {\rm for\ } l \ {\rm half\ odd\ integer} \,. \end{array} \right.
\label{eqFBf}
\eea
Note in particular that we have chosen to define $Y_l$ and $Z_l$ 
alternatingly as the truncated character of the $s$ and $c$ to 
encorporate the different fusion rules for $n$ even or odd.
In terms of a single fugacity $x_s=x_c=x$ the recursion relations
are
\bea
Y_{l+{1\over2}} & \eql & (1-x^2q^{2l}) Z_l + xq^{ {n\over8}+l} 
\chi_{\La_{n-1}} X_l \,,\nonu
Z_{l+{1\over2}} & \eql & (1-x^2q^{2l}) Y_l + xq^{ {n\over8}+l} 
\chi_{\La_{n}} X_l \,,\nonu
X_{l+{1\over2}} & \eql & x^2 q^{2l+{1\over2}}  \chi_{\La_1} X_l +
(1-x^2q^{2l}) \chi_{\La_0} X_{l-{1\over2}} \nonu
&& + x^2 \sum_{i=2}^{n-2} q^{2l + 1 -{i\over2}} 
\left( \prod_{k=0}^{i-2} (1-x^2q^{2l-k}) \right) \left(
\sum_{j=0}^{[i/2]} \chi_{\La_{i-2j}} \right) X_{l - {i-1\over2}} \nonu
&& + xq^{-{n\over8} + l+{1\over2}}  
\left( \prod_{k=0}^{n-3} (1-x^2q^{2l-k}) \right) \chi_{\La_{n-1}} 
Y_{l - {n-3\over2}} \nonu
&& + xq^{-{n\over8} + l+{1\over2}}  
\left( \prod_{k=0}^{n-2} (1-x^2q^{2l-k}) \right) \chi_{\La_n}
Y_{l - {n-2\over2}} \nonu
&& - x^2 \sum_{i=2}^{n} q^{2l - {n+i-4\over2}} 
\left( \prod_{k=0}^{n+i-4} (1-x^2q^{2l-k}) \right)
\left( \sum_{j=0}^{[(n-i)/2]} \chi_{\La_{n-i-2j}} \right) 
X_{l-{n+i-3\over2}} \,.\nonu
&&\label{eqFBg}
\eea
It is trivial to generalize these relations to account for
different fugacities $x_s$ and $x_c$ for the $s$ and $c$ spinons,
respectively.  One simply replaces $x^2$ by either $x_s^2$ ($x_c^2$)
or $x_sx_c$ depending on the fusion rules (\ref{eqFBd}) and (\ref{eqFBe}).  
For examples, see sections 6.4 and 6.6.

After putting $X_l \sim \la_{\rm tot}^l$, 
$Y_l\sim \mu \la_{\rm tot}^{l-{1\over2}}$
and
$Z_l\sim \mu \la_{\rm tot}^{l-{1\over2}}$, putting $q=1$ and specializing 
the characters to the dimensions we find, as in section 6.1, the equation
\be 
(1-\ze)^2 \eql x^2 (1+\ze)^{2n-2}\,, \label{eqFBh}
\ee
where 
\be 
\ze \eql (1-x^2)\la_{\rm tot}^{-{1\over2}}\,,
\qquad \mu \eql {x 2^{n-1} \over 1-\ze}\,.
\label{eqFBi}
\ee
It follows (for $n>1$)
\be
\al \eql 1\,,\qquad n^{\rm max}_{\rm tot}  \eql  4 \,,\label{eqFBj}
\ee
while $\ccft = {n}$, by a computation similar to the one in
(\ref{eqFAj}).

We have the following solution to the recursion 
relations for $\fso(2n)_1$ in terms of the $\fso(2n)$ Milne 
polynomials
\bea
X_{l} & \eql &  q^{ 2 l^2 }   M_{2l\La_1}(x,q^{-1})\,,\nonu
Y_{l} & \eql & q^{ {n\over 8} +   l(2l-1) } 
M_{(2l-1)\La_1+\La_{n-1}}(x,q^{-1})\,,\nonu
Z_{l} & \eql &  q^{ {n\over 8} +  l(2l-1) } 
M_{(2l-1)\La_1+\La_n}(x,q^{-1})\,, \label{eqFBk}
\eea
where $x_a=x^2$ for $a=1,\ldots,n-2$ and $x_a=x$ for $a=n-1,n$.
Explicit expressions for the Milne polynomials entering in (\ref{eqFBk})
at $x=1$ 
in terms of quasi-particle Fock space characters can be given.  Their
forms are similar to those for $\fso(2n+1)$ (cf.\ (\ref{eqFAm})) and will
therefore be omitted.

\subsection{$\sfso(1)_1$ and the Ising model}

Formally, the Ising model can be viewed as the $\fso(2n+1)_1$ WZW
model for $n=0$.  The three primary fields of the Ising model, the
identity $\id$, the energy operator $\psi$ and the spin operator
$\sigma$, correspond to the three `integrable' representations of
$\fso(1)_1$, i.e., the identity, the vector and the spinor
representation, respectively.  The fusion rules, central charge and
conformal dimensions are given by the $\fso(2n+1)_1$ expressions
(\ref{eqFAa}) -- (\ref{eqFAd}) for $n=0$.
In particular, the dimension of the
$\fso(2n+1)$ vector and spinor representation are both $1$ for $n=0$.
Therefore, it is interesting to see if we can find a basis of the
Ising representations in terms of the quasi-particle $\sigma$. 
With the same definition for $X_l$ and $Y_l$ as in (\ref{eqFAe}),
the appropriate recursion  relations for $n=0$ read (cf.\ (\ref{eqFAf}))
\bea
Y_{l+{1\over2}} & \eql & (1-x^2q^{2l}) Y_l + xq^{{1\over16}+l}X_l \,, \nonu
X_{l+{1\over2}} & \eql & (1-x^2q^{2l}) X_{l-{1\over2}} + x^2
q^{2l+{1\over2}} X_l +
x q^{l+{7\over16}} (1-x^2q^{2l})  Y_l \,.
\label{eqFCa}
\eea
The solution to these $\fso(1)_1$ recursion relations behaves,
however, crucially different than the corresponding ones for
$\fso(2n+1)_1$, $n\geq1$.  While the latter are polynomials in
$x$ and $q$ with positive integer coefficients, the solution to
(\ref{eqFCa}) has negative coefficients as well.  In fact, by
construction, we always have $X_l = Y_l = 1$ for $x=q=1$.  As
$l\to\infty$ the negative coefficients get pushed off to infinity,
however, and the solutions $X_l$ and $Y_l$ (for $x=1$) do approach the
correct Ising model characters. We find the following characteristic
equation for $\la(x)$ at $q=1$
\be
\la^{3\over2} - \la - \la^{1\over2} (1-x^2) + (1-x^2)^2 \eql 0\,,
\label{eqFCb}
\ee
which, in terms of $\ze$ defined as in 
(\ref{eqFAh}), can be written as (\ref{eqFAg}) with $n=0$.  The
small $x$ expansion is  
\be
\la(x) = 1 + \sqrt{2} x + \cO(x^2) \,.
\ee

In fact, it is possible to show that (\ref{eqFCa}), for $x=1$, is solved by 
\bea
X_l & \eql & q^{2l^2}    M_{(2l,0)}(q^{-1}) \,,    \nonu
Y_l & \eql & q^{{1\over16}+l(2l-1)} M_{(2l-1,1)}(q^{-1}) \,,\label{eqFCd}
\eea
where $M_{(l,m)}(q)$, for $m=0,1$, are given by (\ref{eqFAm}) and 
(\ref{eqFAn}) for $n=0$.  Explicitly, after simplification,
\bea
M_{(m,0)}(q) & \eql & \sum_{p\geq0} (-1)^p q^{{1\over2}p(p-1)} { (q)_m \over
(q)_{p} (q)_{m-2p} }\,, \nonu
M_{(m,1)}(q) & \eql & \sum_{p\geq0} (-1)^p q^{{1\over2}p(p-1)+(m-p)} 
{ (q)_m \over (q)_{p} (q)_{m-2p} } \,.\label{eqFCe}
\eea
Or, after inverting and redefining the summation variable $p$,
\bea
X_l & \eql & \sum_{p\ {\rm even/odd}} q^{ {1\over2} p^2 } 
{ (q)_{2l} \over (q)_{p} (q)_{l - 
{p\over2} }} \,, \nonu
Y_{l+{1\over2}} & \eql & q^{{1\over16}}
\sum_{p\ {\rm even/odd}} q^{ {1\over2} p(p-1) } 
{ (q)_{2l} \over (q)_{p} (q)_{l - 
{p\over2} }} \,,
\label{eqFCf}
\eea
where the sums are over even $p$ for $l\in\ZZ$ and odd $p$ for
$l\in\ZZ+{1\over2}$. By taking the $l\to\infty$ limit in (\ref{eqFCf}) 
we reproduce well-known fermionic expressions for the Ising model 
characters.  

The finitized Ising characters (\ref{eqFCf}) are different from the 
finitizations obtained from the Majorana fermion description of the 
Ising model (cf.\ eqn.\ (\ref{eqBCa})). 
In fact, one can show that (\ref{eqFCf}) correspond
to an alternative, so-called Bailey or $M$-finitization, of the 
Ising model Virasoro characters \cite{Wa}.  For the Virasoro 
characters $\chi_{(1,s)}(q)$,  the latter take the form
\be
\chi_{(1,s)}^{(l)}(q) \eql q^{h_{(1,s)}}\, 
\sum_j \left( q^{j (12j+4-3s)} \qbin{2l}{l-3j} -
       q^{(3j+1)(4j+s)} \qbin{2l}{l-3j-1 } \right) \,,
\label{eqRCg}
\ee
and we have $X_l = \chi_{(1,1)}^{(l)}$, $X_{l+{1\over2}}= 
\chi_{(1,3)}^{(l)}$ and $Y_{l+{1\over2}}=\chi_{(1,2)}^{(l)}$
for $l\in\ZZ$.
The identity between the expressions (\ref{eqFCf})
and (\ref{eqRCg}) follows from a single iteration of one
of Slater's identities (eqn.\ (A.5) in \cite{Sl}).

As argued above, we expect that the truncated characters are somehow 
related to treating the spin operator $\sigma$ as the fundamental 
quasi-particle.  The precise meaning of this statement has, due to 
the negative coefficients in the truncated characters $X_l$ and $Y_l$, 
so far eluded us.

\subsection{$\sfso(2)_1$}

The $n\to1$ limit of the $\fso(2n)_1$ WZW model can be thought of as 
a $\ccft=1$ model with four representations $\id,v,s$ and $c$ 
with fusion rules as in (\ref{eqFBd}).  
Here the primary fields corresponding to the 
$\id, s$ and $c$ transform in a 1-dimensional representation
of $\fso(2)$ while the primary field for the $v$ transforms as a doublet.  

The appropriate recursion relations for the characters built from the
$s$ and $c$ quasi-particles read (cf.\ (\ref{eqFBg}))
\bea
Y_{l+{1\over2}} & \eql & (1-x_sx_cq^{2l}) Z_l + x_s q^{{1 \over 8}+l}  X_l 
\,,\nonu
Z_{l+{1\over2}} & \eql & (1-x_sx_cq^{2l}) Y_l +  x_cq^{{1 \over 8}+l} X_l
\,,\nonu
X_{l+{1\over2}} & \eql  & (x_s^2+x_c^2)q^{2l+{1\over2}}  X_l +
(1-x_sx_cq^{2l}) X_{l-{1\over2}}  \nonu
&& + q^{{3 \over 8}+l} (1-x_sx_cq^{2l})( x_c Y_l + x_s Z_l ) \,.
\label{eqFDa}
\eea
Here we have introduced separate fugacities $x_s$ and $x_c$ for the 
$s$ and $c$ quasi-particles according to the remarks in section 6.2.

With the asymptotic behavior $X_l \sim \la_{\rm tot}^l$,
$Y_l\sim \mu_s \la_{\rm tot}^{l-{1\over2}}$, and
$Z_l\sim \mu_c \la_{\rm tot}^{l-{1\over2}}$, 
the equation for $\lambda_{\rm tot}$ at $q=1$ becomes
\be
\lambda_{\rm tot}^2 - (x_s^2+x_c^2) \lambda_{\rm tot}^{3 \over 2}
- (1-x_sx_c)(2+x_sx_c) \lambda_{\rm tot} + (1-x_sx_c)^3 \eql 0 \,.
\label{eqFDb}
\ee
Note that for $x_s=x_c=x$ this equation reduces to eqn.\ (\ref{eqFBh})
for $n=1$, with, however, $n^{\rm max}_{\rm tot}=2$.
Equation (\ref{eqFDb}) is the same as that obtained for a system of 
two particles with Haldane statistics and statistical interaction
matrix 
$G = \left( \begin{array}{cc} \scriptstyle{1\over4} & 
\scriptstyle{3\over4} \\
\scriptstyle{3\over4} & \scriptstyle{1\over4} \end{array} \right)$.
Indeed, eqn.\ (\ref{eqBj}) leads to 
\bea
(\lambda_s-1) \left( {\lambda_c \over \lambda_s} \right)^{3\over4}
&=& x_s
\,,\nonu
(\lambda_c-1) \left( {\lambda_s\over \lambda_c} \right)^{3\over4}
&=& x_c  \,,\label{eqFDc}
\eea
which reduces to eqn.\  (\ref{eqFDb}) for $\la_{\rm tot}=\la_s \la_c$.
This result shows that the exclusion statistics of $\fso(2)_1$
spinors are of the type proposed by Haldane.

\subsection{$\sfso(3)_1$}

This case coincides with $\fsu(2)_2$, which has been investigated
elsewhere \cite{BLS2,FrS} (see also section 5.2). The recursion reads
\bea
Y_{l+{1\over2}} & \eql &  (1-x^2q^{2l})Y_l 
+ xq^{ {3\over 16} +l } \chi_{{\bf 2}} X_l
\,, \nonu
X_{l+{1\over2}} & \eql &   x^2 q^{2l+{1\over2}}\chi_{{\bf 3}} X_l +
(1-x^2q^{2l}) X_{l-{1\over2}} \nonu
& & + x q^{{5\over 16} + l}  (1-x^2q^{2l} ) \chi_{{\bf 2}} Y_l \,.
\label{eqFEa}
\eea
where the $\fsu(2)$ characters $\chi_{{\bf 2j+1}}$ are
explicitly given by 
$\chi_{{\bf 2}}=t+t^{-1}$ and $\chi_{{\bf 3}}=t^2+1+t^{-2}$. At $q=1$,
the equation for $\la_{\rm tot}(x)$ reads
\be
\la_{\rm tot}^{3 \over 2} - (1+x^2(t^2+t^{-2})) \la_{\rm tot}
- (1-x^2)(1+x^2) \la_{\rm tot}^{1 \over 2} + (1-x^2)^3 \eql 0 \,.\label{eqFEb}
\ee
Note that for $t=1$ this equation reduces to eqns.\ (\ref{eqFAg}) 
and (\ref{eqFAh}) for $n=1$.

\subsection{$\sfso(4)_1$}

Since $\fso(4)_1\cong \fsu(2)_1 \oplus \fsu(2)_1$ and the $\fso(4)$ spinors
$s$ and $c$ transform as the $(2,0)$ and $(0,2)$ under $\fsu(2)\oplus
\fsu(2)$, respectively, one would expect the solution to the $\fso(4)_1$
recursion relations to exhibit a similar factorization in terms of the
solution to the $\fsu(2)_1$ recursion relations (cf.\ section 2.3).
This is indeed the case as we will now show.

Denoting $\chi_s = \chi_{\La_1}$, $\chi_c=\chi_{\La_2}$, and 
introducing separate fugacities $x_s$ and $x_c$ for the spinors,
the $\fso(4)_1$ recursion relations are (cf.\ (\ref{eqFBg}))
\bea
Y_{l+{1\over2}} & \eql & (1-x_s^2q^{2l}) Z_l + x_sq^{{1 \over 4}+l} 
\chi_{s} X_l \,,\nonu
Z_{l+{1\over2}} & \eql & (1-x_c^2q^{2l}) Y_l + x_c q^{{1 \over 4}+l} 
\chi_{c} X_l \,,\nonu
X_{l+{1\over2}} & \eql &  x_s x_cq^{2l+{1\over2}} \chi_{s} \chi_{c} X_l +
(1-x_s^2q^{2l}) (1-x_c^2q^{2l}) X_{l-{1\over2}} \nonu
&& + q^{l+{1 \over 4}} \left[
   x_s(1-x_c^2q^{2l})\chi_{s} Y_l + x_c(1-x_s^2q^{2l})\chi_{c} Z_l  
   \right] \,. \label{eqFFa}
\eea
In terms of $\fsu(2)_1$ quantities $R^s_l(x_s)$ and 
$R^c_l(x_c)$ satisfying 
\bea
R^s_{l+{1 \over 2}} \eql (1-x_s^2q^{2l})R^s_{l-{1 \over 2}} 
   + x_s q^{l+{1 \over 4}}\chi_{s}R^s_l\,, \label{eqFFb}
\eea
(equivalent to the matrix recursion eqn.\ (\ref{eqBBb}))
and similar for $R^c_l(x_c)$, the recursion (\ref{eqFFa}) is solved by
\bea
X_l \eql R^s_l R^c_l \,, \qquad 
Y_l\eql R^s_l R^c_{l-{1 \over 2}} \,, \qquad 
Z_l\eql R^s_{l-{1 \over 2}} R^c_l \,. \label{eqFFc}
\eea
This factorization expresses the decomposition
$\fso(4)_1 \cong \fsu(2)_1 \oplus \fsu(2)_1$ at the level of
truncated partition sums.

Note that upon specializing the $\fsu(2)$ characters by their dimension,
the recursion (\ref{eqFFb}) is solved by $R^s(x_s)=(1+x_s)^{2l}$.
Thus, for $\fso(4)_1$ and equal chemical potentials for all components, 
the $s$ and $c$ spinons behave as four real fermions.  
Also, note that for $x_s=x_c=x$ the $\fso(4)$ results 
are consistent with (\ref{eqFBh}) and (\ref{eqFBi}).  We conclude that,
despite having to separate $\fso(n)$ for $n$ even or odd, and having
to single out several low $n$ cases, the results for $\la(x)$ are
universal for $n\geq2$.

\subsection{$\sfso(5)_1$}

The case of $\fso(5)_1$ is covered by the analysis in section 6.1.  
In addition to the results mentioned there an explicit formula is
known for the $N$-spinon decomposition. It is given by 
(cf.\ \cite{Yam} for $x=1$)
\be
{\rm ch}_{\la}(x,q) \eql q^{ h(\la)} \ \sum_{\mu=m_1\La_1+m_2\La_2} 
\ x^{2m_1+m_2}
\, {1\over (q)_{m_1}(q)_{m_2} } \ M^{(1)}_{\la\mu}(q)  
  M_\mu(1,q)\,.
\ee
where $\la$ labels the $\fso(5)_1$ sector, i.e.,
$\la=r\La_1 + s \La_2$ ($(r,s)=(0,0)$, $(1,0)$ or $(0,1)$), $h(\la)$ is
given in eqn.\ (\ref{eqFAb}),
and $M_\mu(q)$ is the Milne polynomial of Appendix A.  
An explicit quasi-particle expression for some of the Milne 
polynomials was given in (\ref{eqFAm}).
The fusion rule factor, $M^{(1)}_{\la\mu}(q)$, for which a 
recipe is given in \cite{Yam}, is explicitly given by
\bea
M_{(0,0),(m,n)}^{(1)}(q) & \eql  & \left\{   \begin{array}{ll} 
q^{ {1\over4} (2m^2 + 2mn + n^2) } \sum_{p\ {\rm even}} \ 
q^{ {1\over 2} p^2 } \qbin{{1\over2}n}{p} & \mbox{$m,n$ even} \\
 & \\
q^{ {1\over4} (2m^2 + 2mn + n^2) } \sum_{p\ {\rm odd}} \ 
q^{ {1\over 2} p^2 } \qbin{{1\over2}n}{p} & \mbox{$m$ odd, $n$ even}
\end{array} \right.  \nonu
M_{(1,0),(m,n)}^{(1)}(q) & \eql &  \left\{  \begin{array}{ll}
q^{ {1\over4} (2m^2 + 2mn + n^2-2) } \sum_{p\ {\rm odd}} \ 
q^{ {1\over 2} p^2 } \qbin{{1\over2}n}{p} & \mbox{$m,n$ even} \\
 & \\
q^{ {1\over4} (2m^2 + 2mn + n^2-2) } \sum_{p\ {\rm even}} \ 
q^{ {1\over 2} p^2 } \qbin{{1\over2}n}{p} & \mbox{$m$ odd, $n$ even} 
\end{array} \right. \nonu
M_{(0,1),(m,n)}^{(1)}(q) & \eql &  \left\{  \begin{array}{ll}
q^{ {1\over4} (2m^2 + 2mn + n^2-1) } \sum_{p\ {\rm even}} \ 
q^{ {1\over 2} p(p-1) } \qbin{{1\over2}(n+1)}{p} & \mbox{$m$ even, $n$ 
odd} \\
 & \\
q^{ {1\over4} (2m^2 + 2mn + n^2-1) } \sum_{p\ {\rm even}} \ 
q^{ {1\over 2} p(p-1) } \qbin{{1\over2}(n+1)}{p} & \mbox{$m,n$ odd}
\end{array} \right.  
\eea


\newsection{$\sfsp(2n)_1$ WZW model}

Let $\La_a$, $a=1,\ldots,n$, denote the fundamental weights of $\fsp(2n)$.
The dimension of the irreducible finite dimensional representation 
$L(\La_a)$ with highest weight $\La_a$ of $\fsp(2n)$ is given by 
\be 
{\rm dim}\, L(\La_a) \eql \bin{2n}{a} - \bin{2n}{a-2}\,,
\label{eqGa}
\ee
The affine Lie algebra $\fsp(2n)_1$ has $n+1$ integrable representations
with highest weights $\La_0, \La_1, \ldots, \La_n$ with conformal 
dimension
\be 
h(\La_a) \eql { a (2n+2-a)\over 4(n+2)}\,,\qquad a=0,\ldots,n \,,
\label{eqGb}
\ee
while the central charge is given by 
\be
\ccft \eql {n(2n+1)\over n+2} \,.
\label{eqGc}
\ee
We will take the fundamental quasi-particle for $\fsp(2n)$ to
transform in the $2n$ dimensional irrep $L(\La_1)$ of $\fsp(2n)$.
The relevant $\fsp(2n)_1$ fusion rules are
\bea
\La_1 \times \La_1 & \eql & \La_0 + \La_2 \,,\nonu
\La_1 \times \La_2 & \eql & \La_1 + \La_3 \,,\nonu
 \vdots & & \qquad \vdots \nonu
\La_1 \times \La_{n-1} & \eql & \La_{n-2} + \La_n \,,\nonu
\La_1 \times \La_n & \eql & \La_{n-1} \,.\label{eqGd}
\eea
Denote the truncated characters in the sector $L(\La_a)$, $a=0,\ldots,n$,
by $P^{(a)}_l(x,q)$, $l\in\ZZ/2$.
The recursion relations for $\fsp(2)_1\cong \fsu(2)_1$ and $\fsp(4)_1\cong
\fso(5)_1$ have been worked out in earlier sections.  Their generalization 
to $\fsp(2n)_1$, $n>2$, will be of the following form:
for $l$ integer
\bea
\tP^{(0)}_{l+{1\over2}} & \eql & \tP^{(0)}_l \,,\nonu
\tP^{(1)}_{l+{1\over2}} & \eql & q^{h(\La_1) + l} x 
  \chi_{\La_1} \tP^{(0)}_l +
       (1-q^{2l}x^2) P^{(1)}_l \,,\nonu
\tP^{(i)}_{l+{1\over2}} & \eql & 
  q^{h(\La_i) + il} x^i \chi_{\La_i} \tP^{(0)}_l 
   + \sum_{j=1}^i q^{h(\La_i) - h(\La_j)+ (i-j)l} x^{i-j} (1-q^{2l}x^2)
   \chi_{\La_{i-j}} \tP^{(j)}_l \nonu
  && + \ldots\,, \qquad i\geq2  \,,  \label{eqGe}
\eea
and for $l$ half odd integer
\bea
\tP_{l + {1\over2}}^{(i)}  & \eql &  \sum_{j=i}^{n-1} 
       q^{ h(\La_i) - h(\La_j) + 
  (j-i)(l+{1\over2})} x^{j-i} (1-q^{2l}x^2) \tP_l^{(j)} \nonu &&+
       q^{h(\La_i) - h(\La_n)+ (n-i)(l+{1\over2})} x^{n-i} \tP_l^{(n)}
  + \ldots \,,\qquad i\leq n-2\,, \nonu
\tP_{l + {1\over2}}^{(n-1)}  & \eql & (1-q^{2l}x^2) \tP_l^{(n-1)} +
       q^{h(\La_{n-1} - h(\La_n) + l+{1\over2}} 
  x \chi_{\La_1} \tP_l^{(n)}\,, \nonu
\tP_{l + {1\over2}}^{(n)}  & \eql & \tP_l^{(n)} \,,\label{eqGf}
\eea
where the $\ldots$ stand for additional correction terms of $\cO(x^2)$
and 
such that the representation content of the truncated characters 
at $x=1=q$ is given by 
\be 
P_l^{(a)} \eql \left\{ \begin{array}{cl}
\chi_{\La_{n-a}} (\chi_{\La_n})^{2l-1}\,, & l\ {\rm integer}\,, \\ 
\chi_{\La_{a}} (\chi_{\La_n})^{2l-1}\,, & l\ {\rm  half\ odd\ integer} \,.
\end{array} \right. 
\label{eqGg}
\ee
Unfortunately, we have not been able to find the complete recursion 
relations.  They turn out to be considerably more complicated than the
the other ones analyzed in this paper, in particular non-fundamental
representations enter the recursion relations.  In principle one might
deduce the recursion relations by demanding that they should
be solved by the $\fsp(2n)$ Milne polynomials, i.e.,
\be
P^{(a)}_l \eql \left\{ \begin{array}{cl}
q^{nl^2 - al} M_{\La_{n-a}+(2l-1)\La_n }(x,q^{-1})\,,& l\ {\rm integer}\,,\\
q^{n(l-{1\over2})^2 + a(l-{1\over2}) } 
M_{\La_{a}+(2l-1)\La_n }(x,q^{-1})\,,  & l\ {\rm  half\ odd\ integer}
  \,, \end{array} \right.
\label{eqGh}
\ee
where $x_a=x^a$, $a=1,\ldots,n$.
The recursion relations (\ref{eqGf}) lead to a small $x$-expansion for
$\la_{\rm tot}$ with 
\be
\al \eql 2 \cos \left( {\pi\over n+1} \right) \,,\label{eqGi}
\ee
corresponding to the largest eigenvalue of the fusion matrix 
related to (\ref{eqGd}) (cf.\ (\ref{eqCe})).


\newsection{Further remarks on level-1 WZW models}

The first remark concerns the status of the various claims made in
this paper, in particular on the recursion relations, the fact that
their solutions approach the correct CFT characters in the
$l\to\infty$ limit, their solutions in terms of Milne polynomials for
the level-$1$ affine Lie algebra case and the various quasi-particle
expressions for these Milne polynomials.  While we have been able to
prove isolated cases of some of these claims, in most cases a rigorous
proof is still lacking.  However, all the claims are substantiated by
extensive Mathematica calculations.  This, together with various
consistency checks, such as the computation of the resulting central
charge $\ccft$, sheds little doubt on the validity of all these claims.
 
The second remark concerns our construction of the recursion relations
for the classical affine Lie algebras at level $1$.  This construction
bears a close resemblance to the reproduction scheme of Yangian
representations in \cite{KRb}.  In fact, the quasi-particles that we
use precisely transform in the minimal representations that are used
to generate all Yangian representations through the reproduction
scheme.  Moreover, for $\fsu(n)$ we have seen that the Yangian
$Y(\fsu(n))$ acts on the integrable highest weight modules of
$\fsu(n)_1$, as well as on the truncated Hilbert spaces.  The last
fact could be understood from the fact that the truncated characters,
i.e., the solution $X_l$ of the recursion relation (\ref{eqEAc}), are
precisely the characters of the Haldane-Shastry spin chain of length
$L=nl$ which is known to have exact $Y(\fsu(n))$ symmetry (even for
finite $L$).  While $Y(\fso(n))$ does not act on the integrable
$\fso(n)_1$ modules, the combinations of $\fso(n)$ characters
occurring in the recursion relations (\ref{eqFAf}) and (\ref{eqFBg})
are precisely the `minimal affinizations' on which $Y(\fso(n))$ does
act (cf.\ \cite{KRb}).  Thus, even for $\fso(n)$, the solution of the
recursion relations and hence also the affine characters are actually
virtual characters of $Y(\fso(n))$.\footnote{The explicit
decomposition of the $\fso(n)_1$ characters as virtual $Y(\fso(n))$
characters has apparently been achieved in \cite{KKNpc}.}  The
interpretation in terms of quasi-particles is as follows \cite{BLS2}.
The spinon representation of $\fso(n)$ extends to a representation of
$Y(\fso(n))$ \cite{Dr}, thus we have an action of $Y(\fso(n))$ on the
one-spinon Fock space.  This action extends to the multi-spinon Fock
space by co-multiplication.  However, this multi-spinon Fock space is
bigger than just a direct sum of integrable modules of $\fso(n)_1$.
To get the integrable modules of $\fso(n)_1$, subtractions are needed.
While these subtractions occur in a $Y(\fso(n))$ invariant manner (and
therefore produce virtual $Y(\fso(n))$ characters) the Yangian no
longer acts on the resulting integrable module.


\newsection{Physical applications and final remarks}

Before closing this paper, we like to mention some applications
of the formalism that we developed.

A particularly interesting application concerns the spinor
quasi-particles for the $\fso(5)_1$ WZW model. In a recent
preprint \cite{BS2}, we have explained the relevance of 
these quasi-particles for the description of a particular
model for strongly correlated electrons on a 2-leg ladder.
The model describes itinerant electrons with kinetic
(hopping) term and various interaction constants.
By tuning some of these constants, one can reach a situation
where the model is in a critical $SO(5)$ superspin phase. 
The low-temperature dynamics are then described by the
$\fso(5)_1$ WZW model. Inspecting the quantum numbers of the 
various primary fields, one finds that the $\fso(5)$
vector quasi-particles carry integer spin $S=0,1$ and charge
$0$ or $\pm 2e$. In contrast, the spinor quasi-particles
carry the quantum numbers of a single electron and can
directly be probed in photo-emission type experiments.
This then opens up the possibility that the non-abelian
statistics of the $\fso(5)$ spinor quasi-particles
can, in principle, be observed in experiments.

In a very recent preprint \cite{RR}, Read and Rezayi
have proposed an interesting new class of non-abelian
quantum Hall states. The corresponding edge theories
are conformal field theories, and the quasi-particle
statistics, both in the bulk and at the edge, are
conveniently studied using the formalism developed
here and in \cite{Sc1}. [See \cite{Sc2} for the 
statistics of quasi-particles over the pfaffian
quantum Hall states.] Some first successes of this
approach have already been reported in \cite{RR}.

For further applications to condensed matter systems,
one wants to go beyond the level of equilibrium 
thermodynamics and apply the quasi-particle formalism 
to transport phenomena. Of great importance then are
finite temperature form factors and Green's functions
for the quasi-particles of choice. While this subject
is still being developed \cite{ELS}, it is clear that 
the fractional statistics carried by quasi-particles is 
expressed in some of these finite temperature 
characteristics.

\bigskip\bigskip
\leftline{\large\bf Acknowledgements}\bigskip

We would like to thank the following people for discussions:
L.-H.~Chim, R.A.J.~van Elburg, S.~Guruswamy, A.W.W.~Ludwig, 
B.M.~McCoy, T.~Nakanishi, A.~Schilling and in particular 
S.O.~Warnaar.
P.B.\ is supported by a QEI$\!$I research fellowship from the 
Australian Research Council. The research of K.S.\ is supported
in part by the foundation FOM of the Netherlands.


\appendix
\newsection{Appendix: $q$-deformed tensor products}

Let ${\bfg}$ be a finite dimensional simple Lie algebra of rank
$\ell$.  Let $\La_a$, $\al_a$ and $\al_a^\vee$ ($a=1,\ldots,\ell$) be
the fundamental weights, simple roots and simple co-roots,
respectively, normalized such that $\theta^2=2$ for the longest root
$\theta$.  Following the conventions of \cite{Yam},
for any pair of dominant integral weights $\la$, $\mu$, we
define a polynomial $\wK_{\la\mu}(q)$ by 
\be 
\wK_{\la\mu}(q) \eql \sum_m \
q^{\wc(m)} \prod_{a=1}^\ell \prod_{i=1}^\infty
\qbin{P_i^a(m) + m_i^a}{m_i^a}\,, \label{eqZa}
\ee
where the sum is taken over all nonnegative integers $m_i^a$
($a=1,\ldots,\ell$, $i=1,2,\ldots$), such that
\be
\mu -\la \eql \sum_{a=1}^\ell \left( \sum_{i=1}^\infty im_i^a \right) 
\al_a\,.\label{eqZb}
\ee
Here
\bea
P_i^a(m) & \eql & (\mu,\al_a^\vee) - \sum_{b=1}^\ell \sum_{j=1}^\infty\, 
\Phi_{ij}^{ab} m_j^b \,,\nonu
\Phi_{ij}^{ab} & \eql & 2 { (\al_a,\al_b)\over \al_a^2 \al_b^2 } 
\, {\rm min}(i\al_a^2,j \al_b^2)\,, \label{eqZc}
\eea
and $\wc(m)$ is the cocharge
\be
\wc(m) \eql {\textstyle{1\over2}} \sum_{a,b=1}^\ell \sum_{i,j=1}^\infty\
m_i^a \Phi_{ij}^{ab} m_j^b \,.\label{eqZd}
\ee
We will refer to the polynomial $\wK_{\la\mu}(q)$ as the dual Kostka
polynomial of the Lie algebra $\fg$.  

For $\bfg = \fsl(n)$, the dual Kostka polynomial $\wK_{\la\mu}(q)$ is
related to the conventional Kostka polynomial (see, e.g., \cite{MD})
$K_{\la\mu}(q)$ by (cf.\ \cite{Kir})
\be
\wK_{\la\mu}(q) \eql q^{n(\mu')} K_{\la\mu'}(q^{-1})\,,\label{eqZe}
\ee
where $\mu'$ is the transposition of $\mu$ as a Young diagram, and
\be
n(\mu) \eql \sum_i (i-1)\mu_i = \sum_i  \ \bin{\mu_i'}{2} \,.\label{eqZf}
\ee

Denote by $V_\la$ the
irreducible highest weight module of $\bfg$ with highest weight
$\la$.  Let $W_a$ denote the ``minimal affinization'' of $V_{\La_a}$,
i.e., the minimal irreducible module of the quantum affine algebra
$U_q(\widehat{\bfg})$ (or Yangian $Y(\bfg)$) such that $V_{\La_a}
\subset W_a$.  Let $n_a = (\mu,\al_a^\vee)$, i.e., $\mu= \sum
n_a\La_a$.  From an analysis of the Bethe equations for a $\bfg$
invariant spin chain (cf.\ \cite{KRa,KRb}) it follows that we have the
following tensor product decomposition as $\bfg$ modules 
\be 
W_1^{n_1}
\otimes W_2^{n_2} \otimes \ldots \otimes W_\ell^{n_\ell} ~\cong~
\sum_{\la}\, \wK_{\la\mu}(1) V_\la\,. \label{eqZg}
\ee
It is therefore natural to define the $q$-deformed tensor product as 
\cite{Yam}
\be
[W_1^{n_1} \otimes W_2^{n_2} \otimes \ldots \otimes 
W_\ell^{n_\ell}]_q ~\cong~
\sum_{\la}\, \wK_{\la\mu}(q) V_\la\,. \label{eqZh}
\ee
We associate character valued polynomials $M_\la(x_a,q)$ to the 
$q$-deformed tensor products (\ref{eqZh}) as follows.
Let $\chi_\la$ denote the formal character of the finite dimensional 
irreducible $\fg$ representation $V_\la$ with dominant integral 
weight $\la$, i.e.,
\be
\chi_\la \eql \sum_{\mu\in P(V_\la)} \ e^\mu\,,\label{eqZi}
\ee
where the sum runs over the weights $P(V_\la)$ of $V_\la$.
The character valued polynomial $M_\la(x_a,q)$, where $\la$ is 
a dominant integral weight, is now defined as
\be
M_\la(x_a,q) \eql \sum_{\mu}\ \left(
\prod_a \, x_a^{(\mu,\al_a^\vee)} \right) \, \wK_{\mu\la}(q)\, \chi_\mu\,.
\label{eqZj}
\ee
Because of eqn.\ (\ref{eqZe}), 
the polynomials $M_\la(x_a,q)$ at $x_a=1$ are `dual' to the 
polynomials $Q'_\la(q)$ defined by Milne \cite{Mil} (see
also \cite{Kir}), i.e.,  
\be 
Q'_\la(q) \eql \sum_\mu \ K_{\mu\la}(q) \,\chi_\mu\,.\label{eqZk}
\ee
Therefore we will sometimes 
refer to $M_\la(x_a,q)$ also as a (dual) Milne polynomial.
Note that the polynomials $Q'_\la(q)$ are also sometimes 
refered to as `modified Hall-Littlewood' polynomials.



\end{document}